\documentclass{aa}
\bibpunct{(}{)}{;}{a}{}{,} 
\usepackage{graphicx}
\usepackage[varg]{txfonts}
\defcitealias{Paper1}{Paper I}

\def\l{\ensuremath{\lambda}}

\newcommand{\kms}{km\,s$^{-1}$}

\newcommand{\Halpha} {H$\alpha$}
\newcommand{\Hbeta}  {H$\beta$}
\newcommand{\Hgamma} {H$\gamma$}
\newcommand{\Hdelta} {H$\delta$}

\newcommand{\CII}  {C\,{\sc ii}}
\newcommand{\CIII} {C\,{\sc iii}}

\newcommand{\FeI} {Fe\,{\sc i}}
\newcommand{\FeII} {Fe\,{\sc ii}}
\newcommand{\HeI}  {He\,{\sc i}}
\newcommand{\HeII} {He\,{\sc ii}}

\newcommand{\NIII} {N\,{\sc iii}}

\newcommand{\SiII} {Si\,{\sc ii}}

\usepackage{amstext}

\begin{document}

\title{Voracious vortices in cataclysmic variables}
\subtitle{II. Evidence for the expansion of accretion disc material beyond the Roche lobe of the accretor in HT Cassiopeia
during its 2017 superoutburst}
\titlerunning{Tomographic study of HT~Cas in superoutburst}

\author{V. V. Neustroev\inst{1} \and S. V. Zharikov\inst{2}}
\institute{Space Physics and Astronomy research unit, PO Box 3000, FIN-90014 University of Oulu, Finland. \\
         \email{vitaly@neustroev.net}
         \and
         Instituto de Astronom{\'i}a, Universidad Nacional Aut{\'o}noma de M{\'e}xico,
         Apdo. Postal 877, Ensenada, 22800 Baja California, M{\'e}xico\\
         \email{zhar@astrosen.unam.mx}
           }
\date{Received: 28 August 2019; accepted: 7 August 2020}

\abstract
 {In \citetalias{Paper1} we showed that the accretion disc radius of the dwarf nova
  \object{HT~Cas} in its quiescent state has not changed significantly during many years of observations. It has
  remained consistently large, close to the tidal truncation radius. This result is inconsistent
  with the modern understanding of the evolution of the disc radius through an outburst cycle.}
 {Spectroscopic observations of HT~Cas during its superoutburst offered us an exceptional opportunity
  to compare the properties of the disc of this object in superoutburst and in quiescence.}
 {We obtained a new set of time-resolved spectra of HT~Cas in the middle of its 2017 superoutburst.
  We used Doppler tomography to map emission structures in the system, which we compared with those
  detected during the quiescent state. We used solutions of the restricted three-body problem
  to discuss again the location of emission structures and the disc size of HT~Cas in quiescence.}
 {The superoutburst spectrum is similar in appearance to the quiescent spectra, although the
  strength of most of the emission lines decreased. However, the high-excitation lines significantly
  strengthened in comparison with the Balmer lines. Many lines show a mix of broad emission and
  narrow absorption components. \Halpha\ in superoutburst was much narrower than in quiescence.
  Other emission lines have also narrowed in outburst, but they did not become as narrow as \Halpha.
  Doppler maps of \Halpha\ in quiescence and of the \Hbeta\ and \HeI\ lines in outburst are
  dominated by a bright emission arc at the right side of the tomograms, which is located at and even
  beyond the theoretical truncation limit. However, the bulk of the \Halpha\ emission in outburst has
  significantly lower velocities.}
 {We show that the accretion disc radius of HT~Cas during its superoutburst has become hot but remained
  the same size as it was in quiescence. Instead, we detected cool gas beyond the Roche lobe of the
  white dwarf that may have been expelled from the hot disc during the superoutburst.}


\maketitle

\section{Introduction}

Accretion discs are found in a wide range of astrophysical environments such as active galactic nuclei,
young stellar objects, and interacting binary stars. Most of the accreting white dwarf (WD) systems
known as cataclysmic variables \citep[CVs,][]{Warner95} harbour accretion discs, which play a major role
in their overall behaviour. The discs in dwarf novae, a subclass of CVs, from time to time undergo
outbursts caused by thermal instability, which switches the disc from a low-viscosity to a high-viscosity
regime. For a comprehensive review of the disc instability model (DIM), see \citet{Lasota2001}.
\citet{Hameury19} reviewed the recent updates of the model.

Short-period ($\lesssim$2~hr) dwarf novae of the SU~UMa type show two types of outbursts: normal outbursts
lasting a few days, and superoutbursts that have a slightly larger amplitude and a longer duration of a few
weeks. The defining property of superoutbursts are superhumps, which are low-amplitude modulations whose
period is slightly longer than the orbital period. Superhumps are \emph{\textup{usually}} explained
by the tidal instability of the accretion disc, which grows when the disc expands beyond the 3:1 resonance
radius $R_{\rm 3:1}$. This causes the disc to become eccentric and to precess, which initiates superhumps
\citep{Osaki96}. The increased brightness and duration of superoutbursts compared to normal outbursts are
explained in the context of the thermal-tidal instability model (TTI), which combines the standard DIM with
additional tidal effects such as an enhanced viscous torque that acts when the disc becomes eccentric \citep{Osaki89}.

It is commonly accepted that the largest disc radius is determined by the tidal influence of the
donor star;  the viscous and tidal stresses become comparable at $r_{\rm max}$ and truncate the disc
\citep{Paczynski,PapaloizouPringle77,IchikawaOsaki94}. The 3:1 resonance can only appear in short-period
systems where the ratio $q \equiv M_{\rm 2}/M_{\rm 1}$ of the masses of donor and WD is low enough,
$\lesssim$0.25--0.33 \citep{WhitehurstKing91}. In such CVs,
$R_{\rm 3:1}$ is lower than $r_{\rm max}$. While the responsibility of the tidal instability for the
superhump phenomenon is still under debate \citep{Bisikalo04,HameuryLasota05}, there is general
consensus that during outbursts, the accretion disc expands and then contracts with time
\citep{Lasota2001,Osaki05,HameuryLasota05}.

\begin{table*}
\caption{Log of spectroscopic observations of HT Cas}
\label{ObsTab}
\centering
\begin{tabular}{clcccc}
\hline\hline
 HJD Start & Telescope/      &  \l~range        & Exp.Time & Number   & Duration \\
 2450000+  & Instrument      &     (\AA)        &  (s)     & of exps. &  (h) \\
\hline
 7771.370  & NOT / ALFOSC    &  4410--6960      &  120     & 50       & 1.83 \\
 7771.449  & NOT / ALFOSC    &  3650--7110      &  200     &  1       & 0.06 \\
 7771.639  & 2.1~m / B\&Ch   &  4080--7560      &  600     &  1       & 0.17 \\
 7772.615  & 2.1~m / B\&Ch   &  4080--7560      &  600     &  1       & 0.17 \\

\hline
\hline
\end{tabular}
\end{table*}

However, in our earlier paper, we showed \citep[hereinafter referred to as \citetalias{Paper1}]{Paper1}
that the disc radius in the dwarf nova of the SU~UMa-type \object{HT~Cas} has not changed significantly
during many years of observations and remained consistently large, close to $r_{\rm max}$. Multi-epoch,
time-resolved spectroscopic observations from \citetalias{Paper1} were obtained between 1986 and 2005
in quiescence, interrupted by several normal outbursts. This result is not consistent with the modern
understanding of the evolution of the accretion disc through an outburst cycle, as described above. We also
note that \citetalias{Paper1} was mostly dedicated to studying the properties of emission structures in
the system, of which the dominated source is the extended emission region at the leading side of the
disc, opposite to the location of the hotspot from the area of interaction between the gas stream and
the disc. This puzzling feature was detected in Doppler maps of many CVs, but its origin still remains
unclear. In \citetalias{Paper1} we found that the leading side bright spot is always observed at the
very edge of the disc.

\begin{figure}[b]
\resizebox{\hsize}{!}{\includegraphics{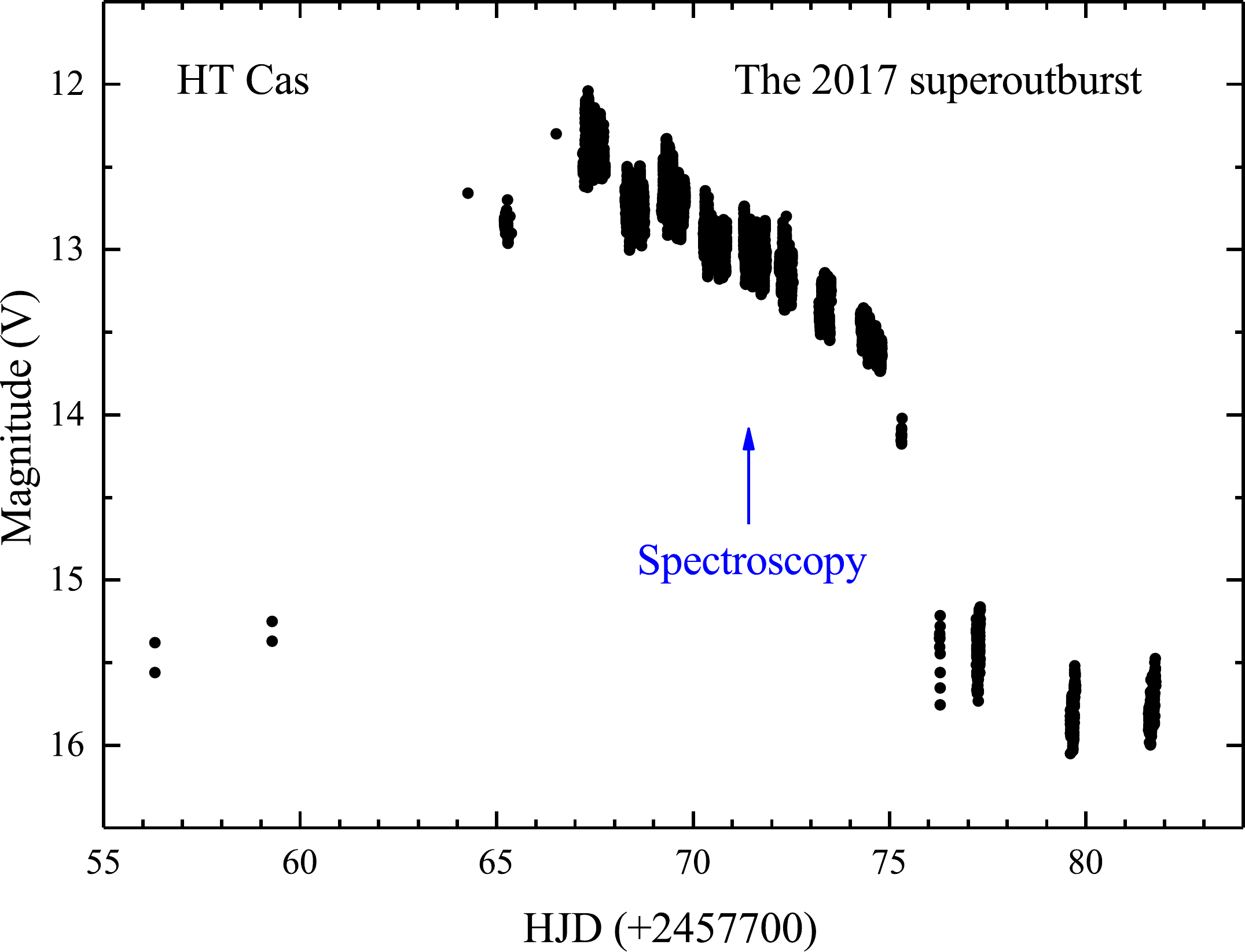}}
\caption{AAVSO light curve of HT Cas during the 2017 superoutburst (only out-of-eclipse observations
         are shown).}
\label{Fig:AAVSO}
\end{figure}

At the beginning of 2017, HT~Cas experienced a very rare superoutburst \citep[the previous superoutburst was
observed in 2010,][]{Kato2012} during which strong superhumps were detected (Enrique de Miguel,
vsnet–alert 20570). This event offered us an exceptional opportunity to provide a crucial test for
the theory. In the middle of the superoutburst (see
Fig.~\ref{Fig:AAVSO}), we obtained a new set of time-resolved spectra of HT~Cas, which was analysed
in a similar manner as our previous observations in quiescence \citepalias{Paper1}. Here we present
a comparable analysis of the properties of the accretion disc and of its emission structures
during its high and low states. We show the most striking result: the accretion disc radius did not
change in the superoutburst because it was already restricted by the tidal limit. Instead, we detected
cool gas beyond the Roche lobe of the WD that might have been expelled from the hot accretion disc.

\begin{figure*}
  \resizebox{\hsize}{!}{\includegraphics[angle=90]{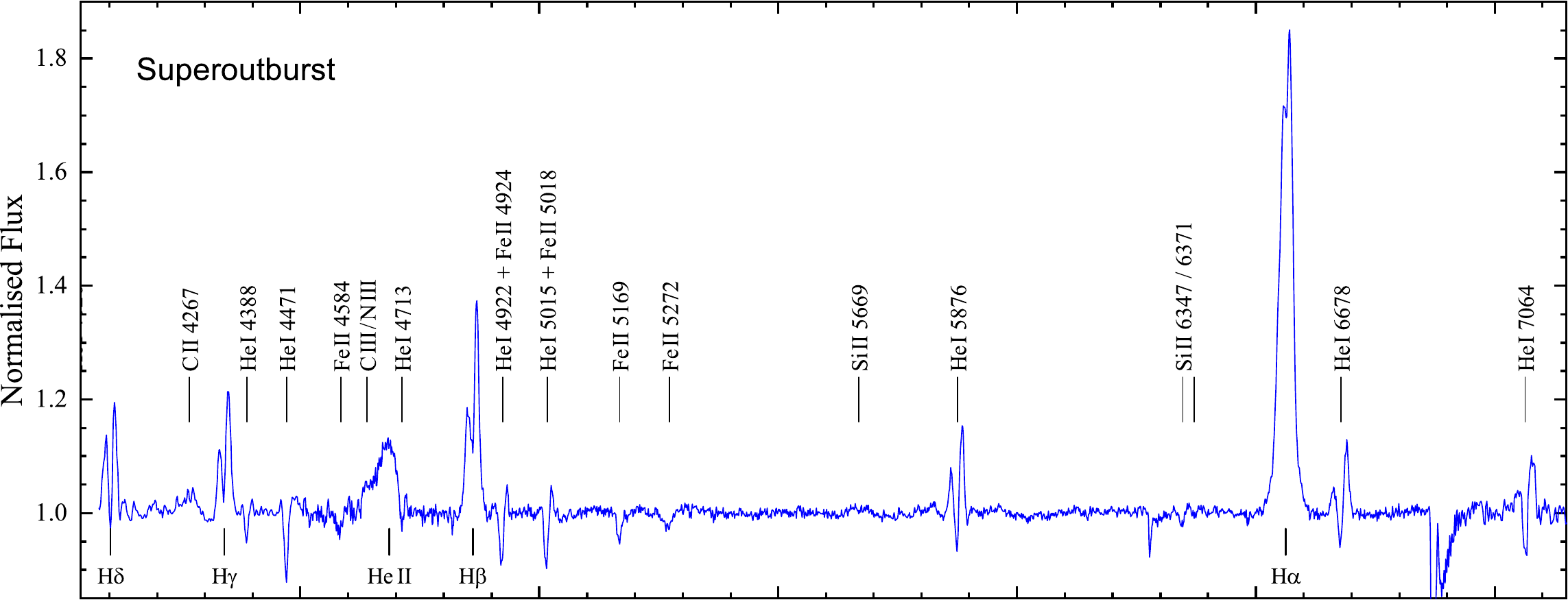}}
  \resizebox{\hsize}{!}{\includegraphics{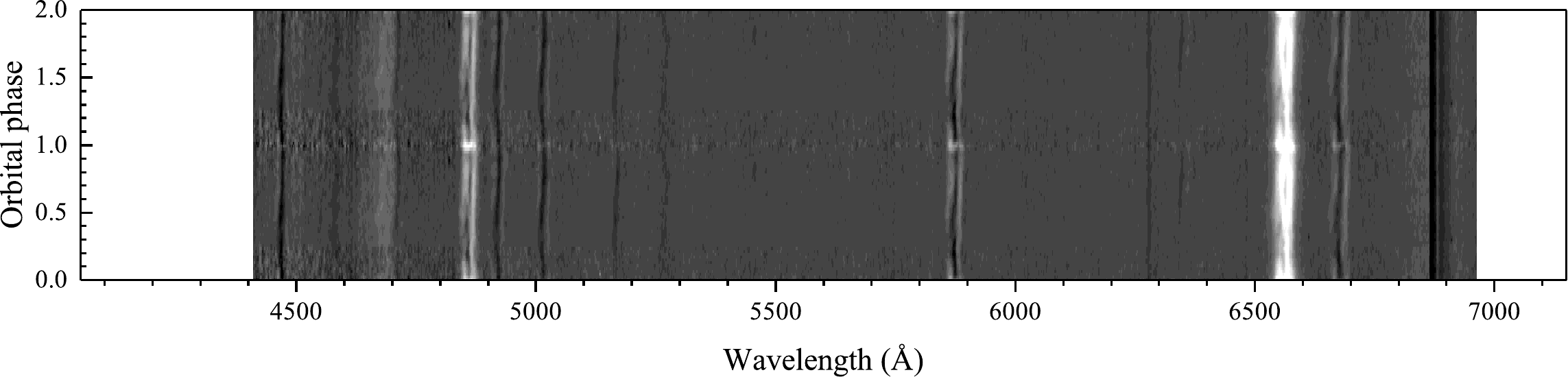}}
  \caption{Averaged and continuum-normalised (top panel) and trailed (bottom panel) spectra of HT~Cas
           during its 2017 superoutburst.
           In the trailed spectrum, displayed twice for clarity, most of the lines consist of a mixture
           of absorption and emission components. White indicates emission (shown on a linear scale).}
  \label{Fig:avespec}
\end{figure*}

\section{Observations and data reduction}
\label{Sec:Obs}

The spectroscopic observations of HT~Cas were performed on 2017 January 17 with the Andalucia Faint
Object Spectrograph and Camera (ALFOSC) mounted at the 2.5 m Nordic Optical Telescope (NOT) in the
Observatorio de Roque de los Muchachos (ORM, La Palma, Spain). The data were obtained under perfect
weather conditions with seeing 0.7--0.9\arcsec, allowing us to use a narrow slit of 0.5\arcsec.
The observations were taken with grism 19 in the wavelength range of 4410--6960 {\AA} with a
dispersion of 1.2~{\AA} pixel$^{-1}$ and a corresponding spectral resolution of 2.9~\AA. A total
of 50 spectra with 120 sec individual exposures were obtained, covering one orbital period of the
system. He-Ne lamp exposures were taken before, in the middle, and after the observations of the
target for wavelength calibration, which is accurate within an rms of 0.040 \AA. An accurate
wavelength scale for each object spectrum was established through interpolation between the nearest
arcs in time. In addition, to extend the wavelength coverage, we also took
one spectrum in the wavelength range of 3650--7110 {\AA} using grism 7. Because no spectroscopic
standards were observed when the spectra of HT~Cas were taken, the latter were first normalised
to the continuum (Figure~\ref{Fig:avespec}).

$\text{About }$4.5 hr after the end of the NOT observations, we obtained an out-of-eclipse spectrum
of HT~Cas with the Boller \& Chivens spectrograph (B\&Ch) attached to the 2.1 m telescope of the Observatorio
Astronomico Nacional (OAN SPM) in Mexico (the same instrument as we used in \citetalias{Paper1}).
In the following night, we obtained another spectrum. The two spectra were flux calibrated with the
standard star Feige34, the spectra of which were obtained in the same nights. The continuum of HT~Cas
in both spectra can be well described by a power law with slightly different indices: $-$2.85 and
$-$2.60 for the first and second nights, respectively. This difference can be due to either a colour
evolution of the object during an outburst \citep[see e.g.][]{NeustroevBZ,NeustroevSSS} or an orbital
colour variation. In the following, we adopt the power index of $-$2.85 and rescale the NOT spectra
to this slope and to match simultaneous optical photometry\footnote{We used the public data from the
AAVSO International Database, http://www.aavso.org/data-download}.
The log of spectroscopic observations is presented in Table \ref{ObsTab}.

\section{Eclipse ephemerides and orbital phases}

Eclipse ephemerides for HT~Cas have been derived many times in the past \citep[e.g.][]{Ultracam,Borges2008}.
However, they have accumulated a significant error over time, which resulted in a large discrepancy between the
predicted and observed eclipse times. Because the time resolution of our spectroscopy is not sufficient
to define an accurate zero-phase, we also used the AAVSO photometric observations of HT~Cas during its
superoutburst \citep{AAVSO}.
Thus, we extracted 47 times of light minima between JD 2457765 and 2457775.
A linear least-squares fit to these times gives the following orbital ephemeris of the light-minima:
\begin{equation}
HJD_{min} = 245\,7765.23997(3)+0.07364720309 \cdot E.
\end{equation}
In this calculations, we \emph{\textup{adopted}} the value of the orbital period $P_{orb}$=0.07364720309~d
from \citet{Ultracam}. The obtained ephemerides were used to calculate the orbital phases of the spectra.

\section{Data analysis}

The data analysis in this paper is mostly performed in a similar manner to that described in \citetalias{Paper1},
and we also used the same plot templates for the figures. Thus, we recommend consulting \citetalias{Paper1}
for technical details of the analysis and comparing the figures in this work with those of \citetalias{Paper1}.
Here we also use the same system parameters of HT~Cas taken from \citet{Horne91}: $M_1 = 0.61\pm0.04M_\odot$,
$M_2=0.09\pm0.02M_\odot$, $q=0.15\pm0.03$, and $i=81.0\pm1.0$\degr.

\subsection{Averaged and trailed spectra}
\label{Sec:Decrement}

Figure~\ref{Fig:avespec} (top panel) shows the averaged and continuum-normalised out-of-eclipse spectrum,
while the measured parameters of the most prominent spectral lines are presented in Table~\ref{Tab:LineParam}.
The superoutburst spectrum is quite similar in appearance to those observed in quiescence (see figure 2
in \citetalias{Paper1}). This is not typical of dwarf novae in outburst; the emission lines
during outbursts are usually replaced by broad absorption troughs with weak emission cores
\citep{NeustroevBZ,NeustroevSSS,GW_Lib,V1838Aql}.
On the other hand, there are quite many examples of CV outbursts during which the spectra still show
strong emission lines \citep{MoralesRueda2002,NeustroevTCP1}. Thus, the case of HT\,Cas is not unique.
We acknowledge the fact, however, that our data cover only a fraction (two days) of the superoutburst
during which a complex spectral evolution was possible.

Despite the similarity of spectra between superoutburst and quiescence, there are at least a few obvious
differences. As in quiescence, the outburst spectrum shows double-peaked emission lines of the Balmer series
and \HeI. However, the relative (to the continuum) intensities and the equivalent widths (EW) of these lines
are significantly lowered (compare Table~\ref{Tab:LineParam} and table 3 from \citetalias{Paper1}).
The outburst spectrum shows a notable strengthening compared to the Balmer lines of
high-excitation emission lines such as \HeII\ 4686, \CIII/\NIII\ 4640--4650, and \CII\ 4267. The
EW of \HeII\ 4686 is comparable in the quiescence and outburst spectra.
The double-peaked emission lines of \SiII\ 5669, 6347, and
6371 are clearly detected, although they are rarely observed in CVs. We also observe many absorption lines
of \FeII\ (e.g. $\lambda\lambda$4549, 4584, 5169, 5182, 5235, and 5272~\AA), and possibly a few of \FeI\
($\lambda$4518~\AA).

\begin{figure}
  \centering
  \includegraphics[width=7.9cm,angle=0]{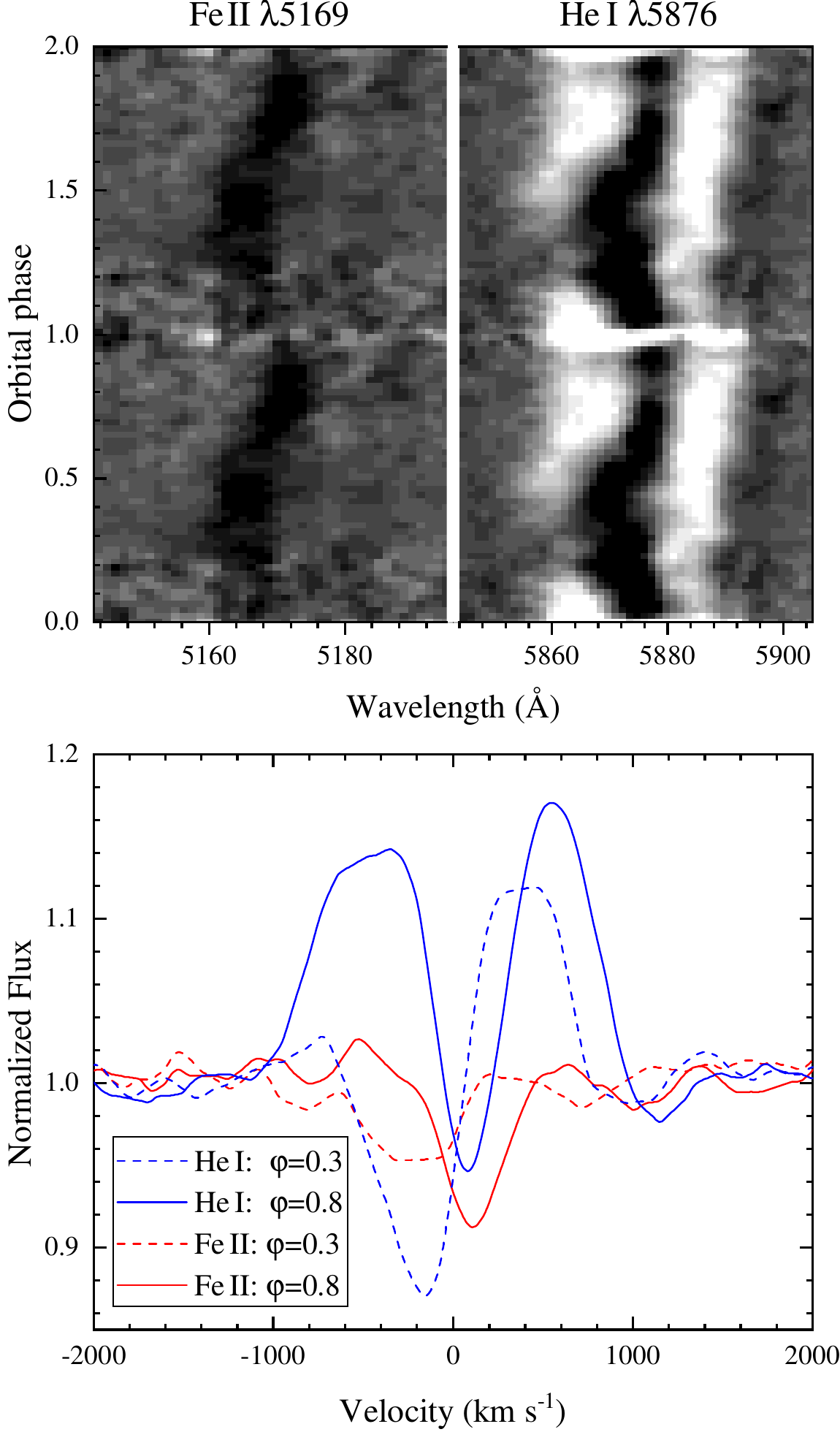}
  \caption{Trailed spectra (top panel) and averaged profiles (bottom panel) of the \FeII\ 5169 and
           \HeI\ 5876 lines. The spectra are averaged around the orbital phases 0.3$\pm$0.1 and 0.8$\pm$0.1.
           Trailed spectra are shown on a linear scale (white indicates emission).}
  \label{Fig:Absorptions}
  \vspace{5mm}
  \centering
  \includegraphics[width=8cm,angle=0]{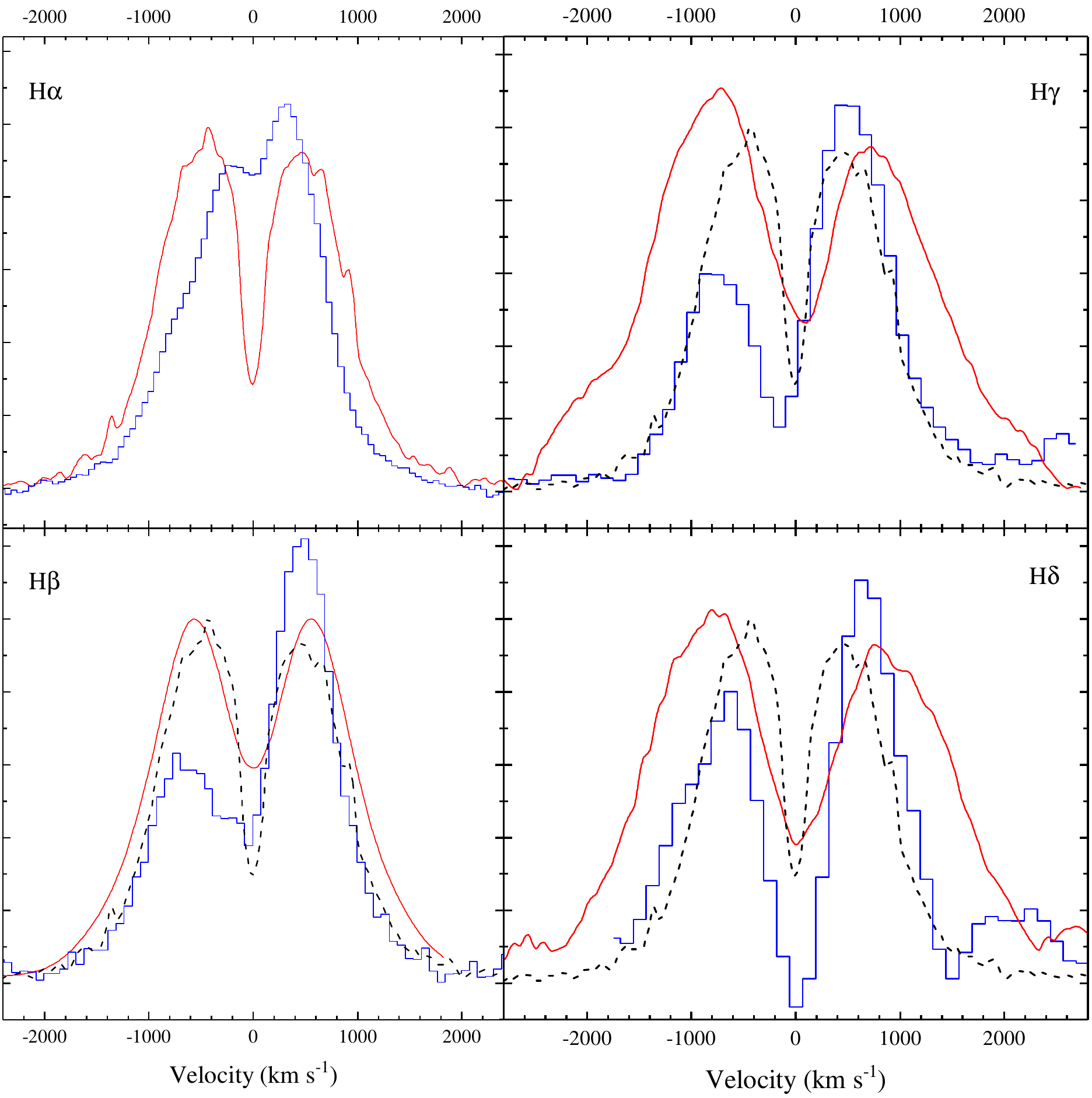}
  \caption{Averaged profiles of the Balmer lines during the 2017 superoutburst (blue)
  and in quiescence (2005), shown in red. \Hbeta, \Hgamma, and \Hdelta\ are also compared
  with the \Halpha\ quiescence profile shown by the dashed black line.}
  \label{Fig:LineProfiles}
 \end{figure}

In the phase-resolved trailed spectra (Fig.~\ref{Fig:avespec}, bottom panel), many lines
show a mixture of broad emission and \emph{\textup{narrow}} absorption components. In the \HeI\ and possibly \SiII\
lines, an absorption core between emission peaks goes below the continuum. The iron
lines seem to exhibit the absorption core alone, or their emission components are very weak. In the \Halpha\ and
\Hbeta\ lines the absorption component is undetectable in the averaged spectra (the central valley between
emission peaks is even more shallow than in quiescence spectra, although it becomes stronger in the higher
order Balmer lines), but a sign of absorption is visible in the trailed spectra. The absorption component
of different lines varies in phase with each other and with a similar, relatively small radial velocity
amplitude. It remains inside the emission peaks and becomes broader at orbital phase $\sim$0.3, when the lines
are blueshifted most strongly. In emission lines, it gives the impression that during these phases, the blue peak
is shadowed, becoming weaker (Fig.~\ref{Fig:Absorptions}). In mid-eclipse, the absorption components of
the Balmer and \HeI\ lines almost disappeared or even reversed into emission.

One of the most important spectral changes between quiescence and the superoutburst is a significant
narrowing of the \Halpha\ profile in outburst (Fig.~\ref{Fig:LineProfiles}). It is not only evident
from the peak-to-peak separation, which decreased from $\sim$1100 to 670 \kms, but also from the full width
at half maximum (FWHM), which also decreased from $\sim$2000 to 1450 \kms\ (Table~\ref{Tab:LineParam}).
It is interesting that most of other emission lines also narrowed (the \Hbeta, \Hgamma, and \Hdelta\
lines are shown in Fig.~\ref{Fig:LineProfiles}), but they did not become as narrow as the \Halpha\
line. In contrast, their width (both the FWHM and peak-to-peak separation) became very similar to each
other and close to the \Halpha\ width in \emph{\textup{quiescence}}. Because the dominant broadening mechanism
of double-peaked emission lines is the Doppler shift due to Keplerian rotation of the accretion disc
around the accretor \citep{Smak1969,Smak1981,HorneMarsh86}, the observed narrowing of all emission lines
suggests their origin during the outburst from a more extended area than in quiescence. We discuss this
in detail in Sections~\ref{Sec:Evidence4beyond} and \ref{Sec:Evidence4Ejection}.

\begin{table}
\caption{Parameters of the most prominent lines in the mean spectrum}
\label{Tab:LineParam}
\centering
\begin{tabular}{llcc}
\hline\hline
Spectral            & Relative   & EW    & Peak-to-peak \\
 line               & intensity$^{c}$  & (\AA) &(\kms) \\
\hline
 \Halpha            &  1.80/--   & -27.8 &  670  \\
 \Hbeta             &  1.28/--   &  -8.0 & 1170  \\
 \Hgamma            &  1.16/--   &  -4.9 & 1060  \\
 \Hdelta$^{a}$      &  1.15/--   &  -3.0 & 1250  \\
 \HeI\ 4471$^{a,b}$ &  1.04/0.85 &  +0.9 & 1300  \\
 \HeI\ 4713$^{a,b}$ &  1.02/0.97 &   --  &  --   \\
 \HeI\ 4922$^{a}$   &  1.03/0.90 &  -0.2 & 1560  \\
 \HeI\ 5015$^{a}$   &  1.03/0.90 &  -0.2 & 1250  \\
 \HeI\ 5876$^{a}$   &  1.12/0.93 &  -2.0 & 1190  \\
 \HeI\ 6678$^{a}$   &  1.07/0.94 &  -1.6 & 1200  \\
 \HeII\ 4686$^{b}$  &  1.12/--   &  -4.7 &  --   \\
 \CIII/\NIII$^{b}$  &  1.05/--   &  -1.8 &  --   \\
 \FeII\ 5169$^{a}$  &    --/0.94 &  +0.6 &  --   \\
 \SiII\ 5669        &  1.01/--   &  -0.7 & 1300: \\
 \SiII\ 6347$^{a,b}$&  1.01/0.98 &  -0.2 & 1350: \\
 \SiII\ 6371$^{b}$  &  1.01/0.99 &  -0.4 & 1200: \\
\hline
\end{tabular}
\tablefoot{\textit{a} -- a strong absorption core is present; \textit{b} -- blend;
  \textit{c} -- intensity relative to the continuum intensities is shown for emission and
  absorption components, if present.
         }
\end{table}

\begin{figure}
  \centering
  \includegraphics[width=8cm,angle=0]{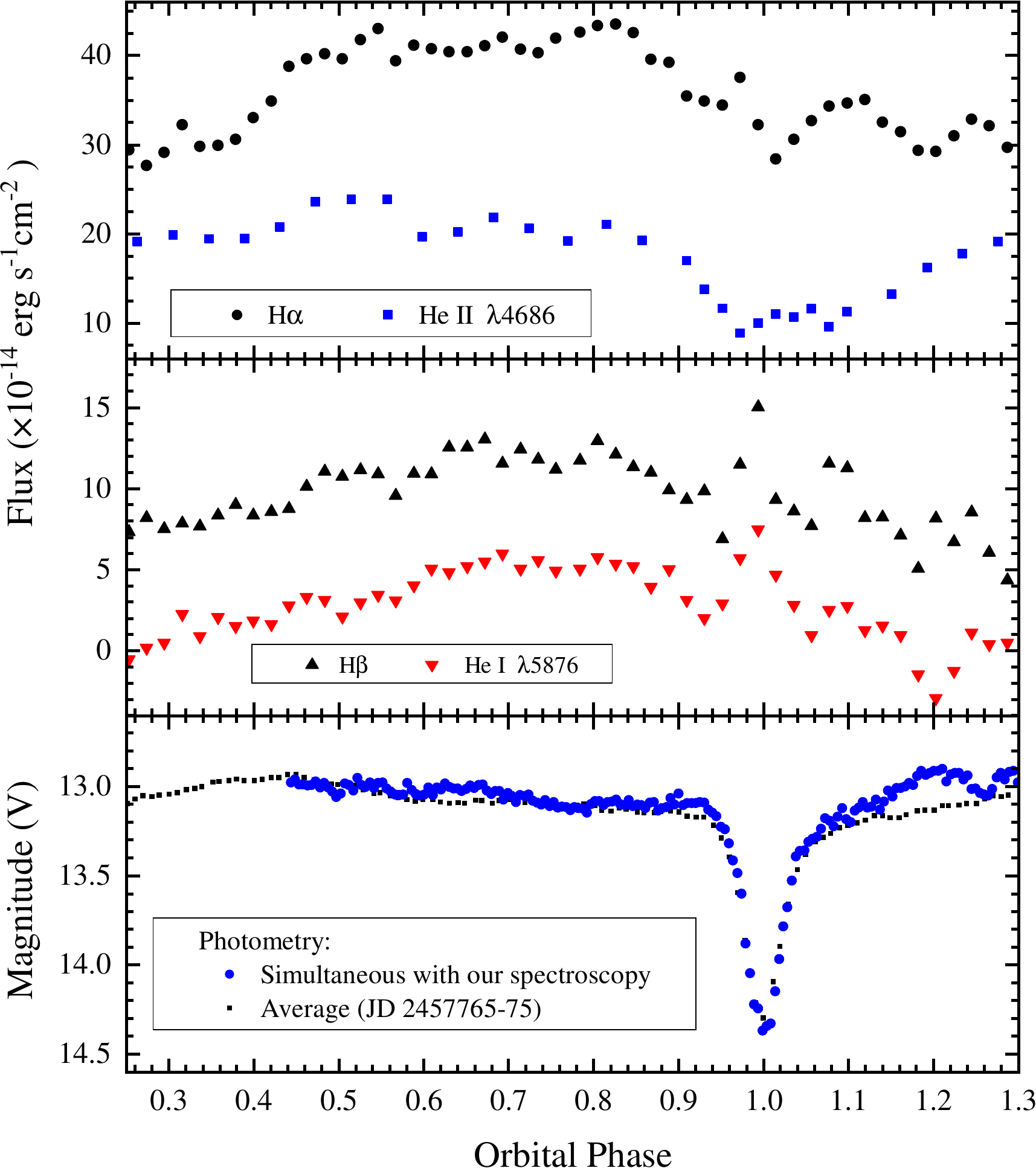}
  \caption{Light curves computed for the \Halpha, \HeII\ \l4686, \Hbeta,\ and \HeI\ \l5876 lines (top
           and middle panels). The bottom panel shows the AAVSO light curves. The black
           squares represent the phase-averaged data obtained between JD\,2457765 and 2457775, and
           the blue circles show the light curve of these data at the time of our spectroscopic
           observations.}
  \label{Fig:lightcurves}
\end{figure}

\begin{figure*}
  \centering
  \includegraphics[width=17cm,angle=0]{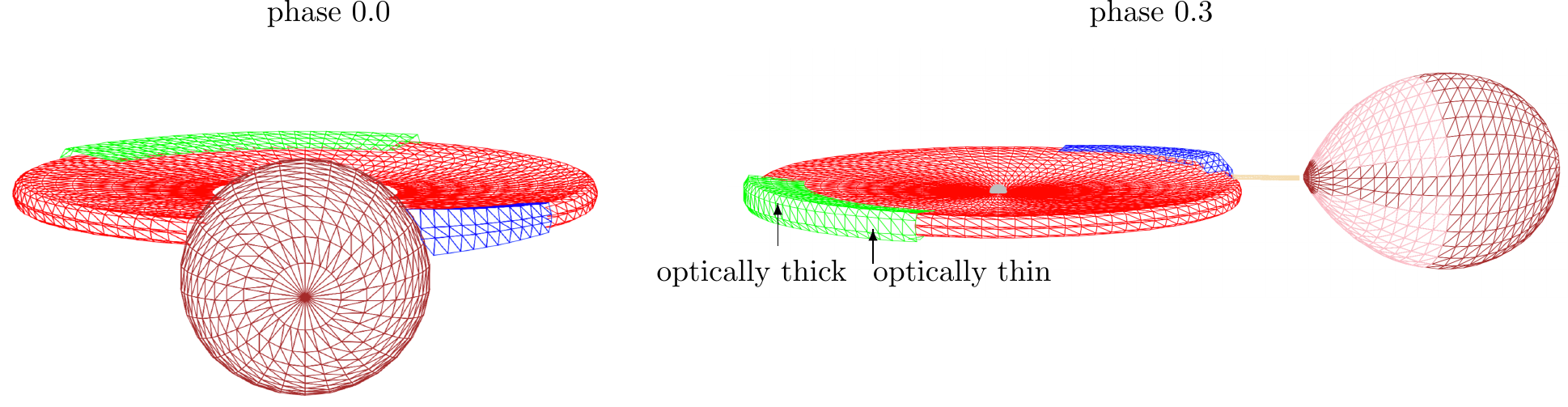}
  \caption{Schematic representation of the HT~Cas geometry in the middle of eclipse (left, orbital
           phase 0.0) and at orbital phase 0.3 (right), when the depression of the blue peak of emission
           lines is observed.}
  \label{Fig:HTCas}
\end{figure*}

\subsection{Light curves}

The flux-calibrated spectra were used to compute light curves for the emission lines \Halpha, \Hbeta,
\HeI\ \l5876, and \HeII\ \l4686. Similarly to \citetalias{Paper1}, we computed the light curves by summing
the continuum-subtracted flux inside of $\pm$2700 \kms\ window centred at the emission line wavelengths.
These light curves are shown in Fig.~\ref{Fig:lightcurves}. In the bottom panel of
Fig.~\ref{Fig:lightcurves} we show two AAVSO light curves, the phase-averaged one obtained between
JD\,2457765 and 2457775, and the light curve obtained during our spectroscopic observations.

By comparing the emission line curves with figure 3 from \citetalias{Paper1}, we conclude that
the character of the variability has changed significantly. In superoutburst it has shown single-wave
modulations rather than the double wave seen in quiescence. However, the most striking difference is the
increase in fluxes (not only in EWs) of some of the emission lines in mid-eclipse. This effect is
not seen in high-excitation lines and is hardly visible in \Halpha,\ but is clearly detected in \Hbeta\ and
very strong in the \HeI\ lines. It can be explained by assuming
that the absorption line components were produced in the innermost parts of the optically thick accretion
disc, which are eclipsed by the donor, whereas the emission line region is located more outward in the
disc and/or possibly above the orbital plane. Figure~\ref{Fig:HTCas} (left-hand panel) schematically
demonstrates the system
geometry in the middle of the eclipse. The inner disc with its absorption spectrum is eclipsed around
mid-eclipse phases, while the emission layer can still be seen. In this case, it becomes the main contributor
to the spectrum, giving the raise to the total line flux. The more pronounced the absorption component,
the stronger the flux increase in mid-eclipse.

\subsection{Doppler tomography}

\begin{figure*}
        \resizebox{\hsize}{!}{\includegraphics{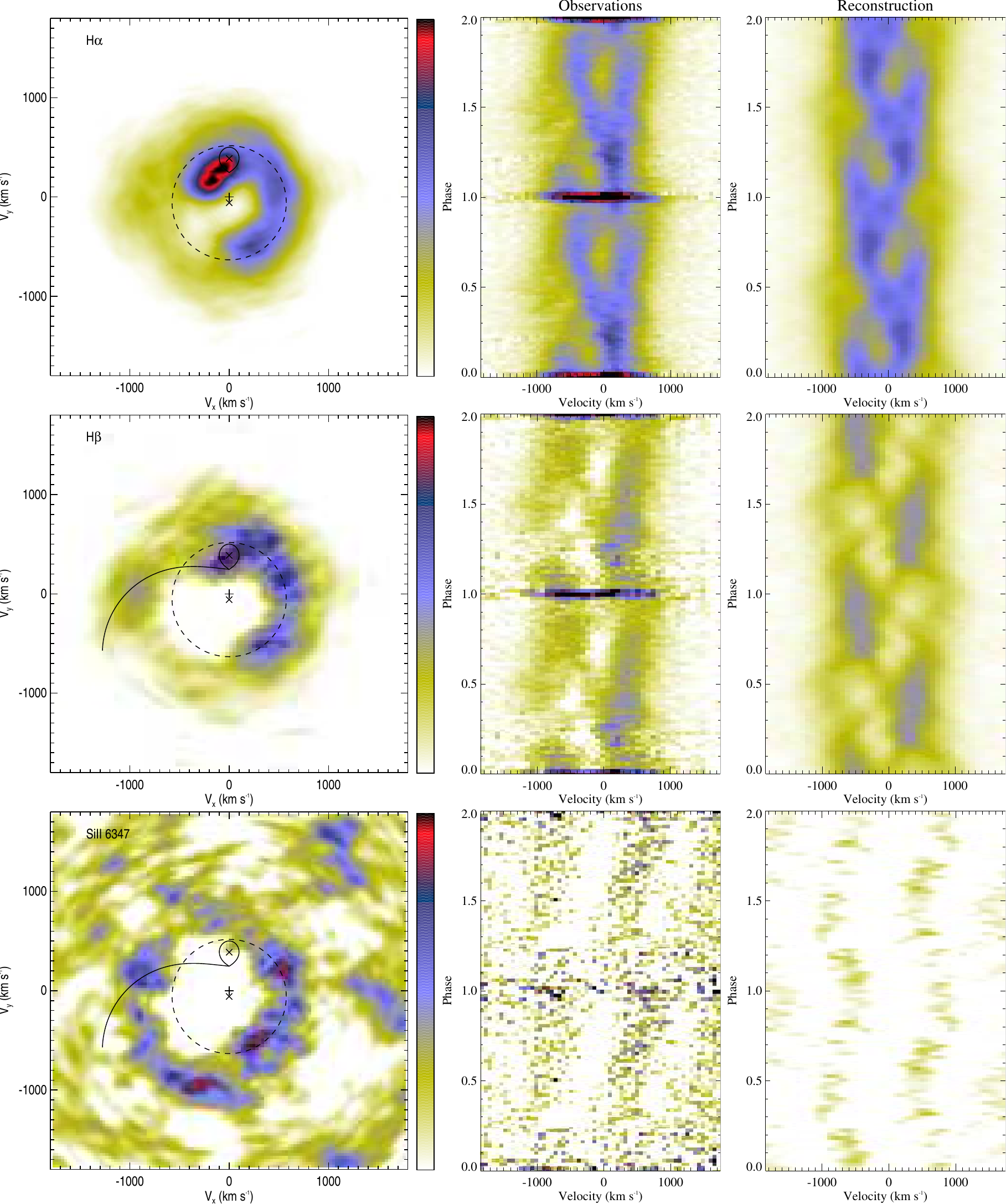}}

    \caption{Doppler maps and corresponding observed and reconstructed trailed spectra of the \Halpha,
             \Hbeta, and \SiII\ 6347 emission lines.
             The position of the WD (lower cross), the centre of
             mass of the binary (middle cross), the Roche lobe of the secondary star (upper bubble
             with the cross), and the predicted trajectory of the gas stream in the form of the curve are marked.
             The circle shows the tidal truncation radius $r_{\rm max}$ of the accretion disc,
             assuming a circular Keplerian flow. The colour bars indicate normalised flux on a linear scale.
             }
    \label{Fig:dopmaps1}
\end{figure*}

\begin{figure*}
    \resizebox{\hsize}{!}{\includegraphics{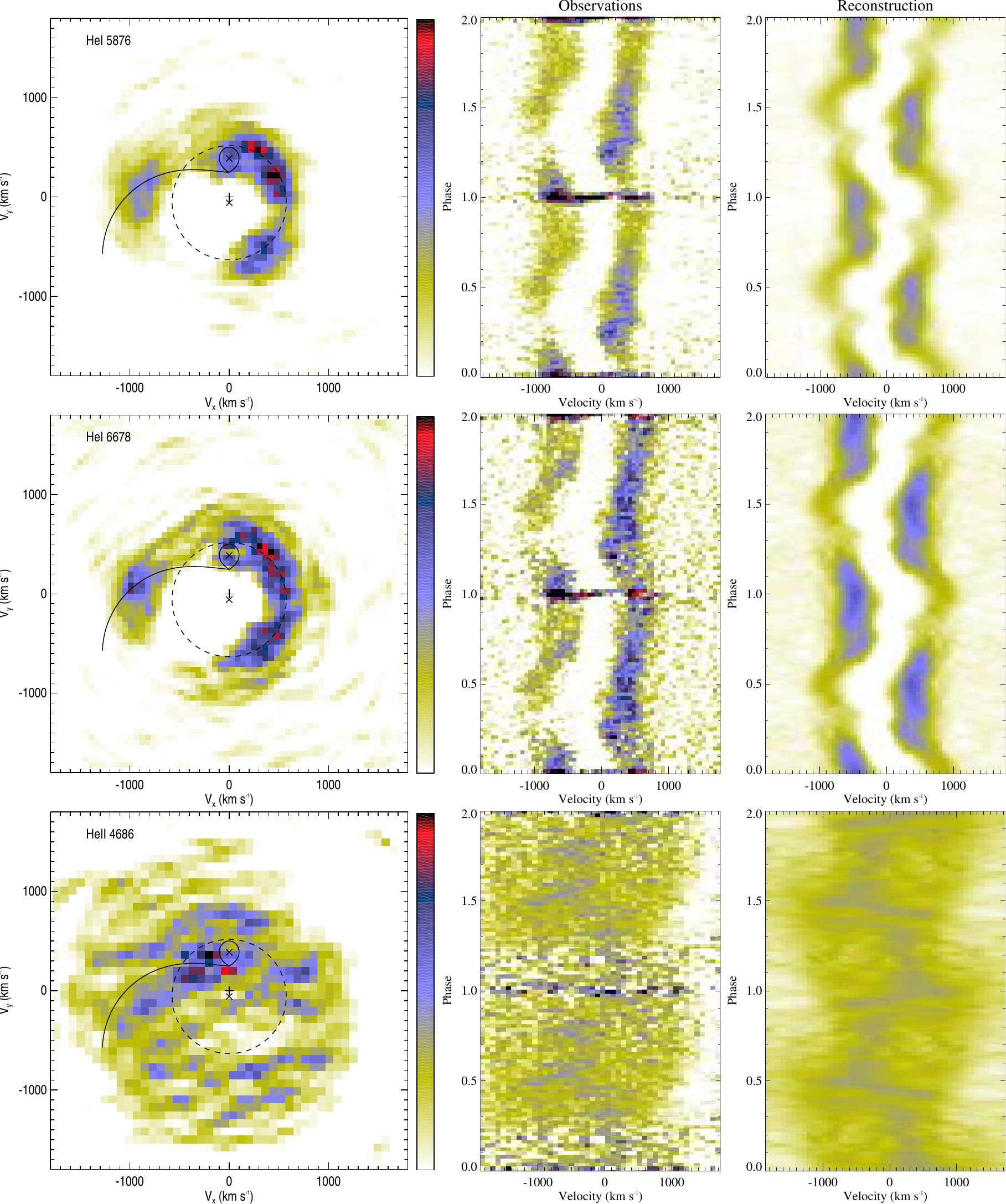}}
    \caption{Doppler maps and corresponding observed and reconstructed trailed spectra of the \HeI\ 5876
             and 6678 and the \HeII\ 4686 emission lines.
             The position of the WD (lower cross), the centre of
             mass of the binary (middle cross), the Roche lobe of the secondary star (upper bubble
             with the cross), and the predicted trajectory of the gas stream in the form of the curve are marked.
             The circle shows the tidal truncation radius $r_{\rm max}$ of the accretion disc,
             assuming a circular Keplerian flow. The colour bars indicate normalised flux on a linear scale.
             }
    \label{Fig:dopmaps2}
\end{figure*}

\begin{figure*}
    \resizebox{\hsize}{!}{\includegraphics{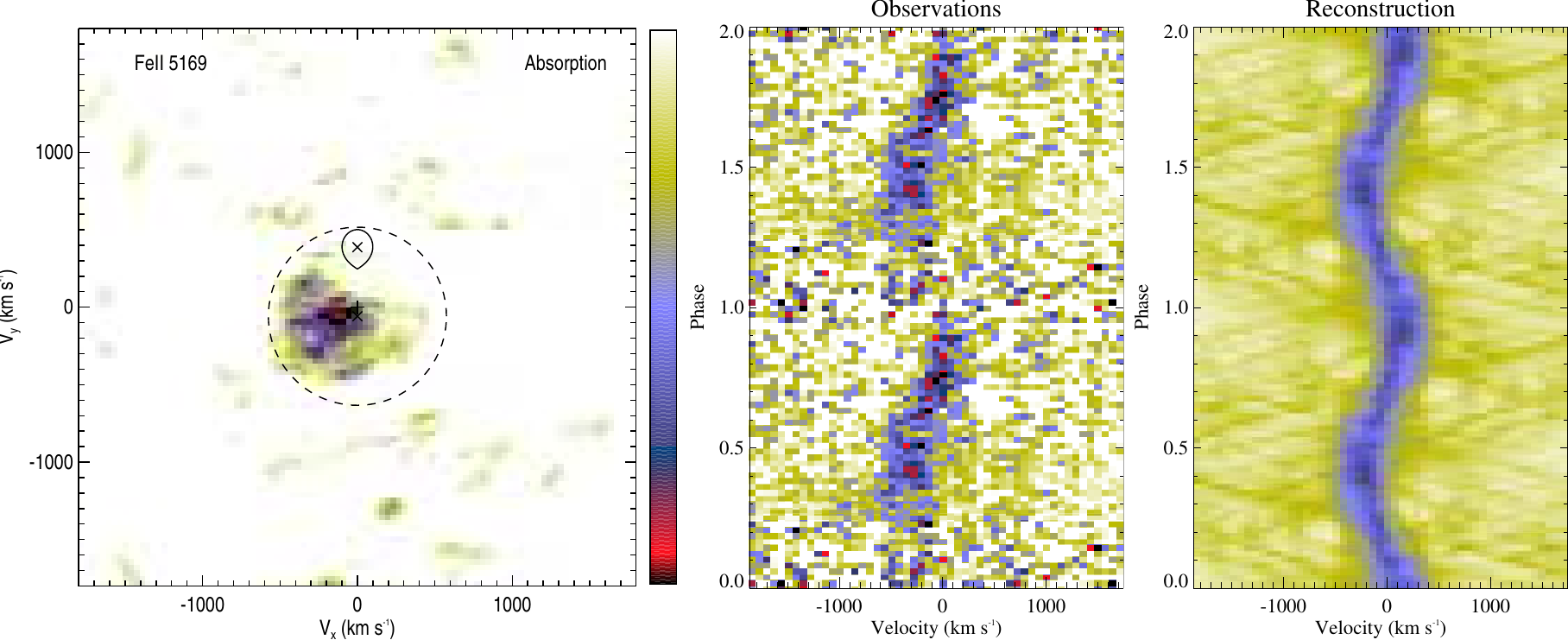}}
    \caption{Doppler map and corresponding observed and reconstructed trailed spectra of the \FeII\ 5169
             absorption line. The colour bars indicate normalised flux on a linear scale.
             }
    \label{Fig:dopmapAbs}
\end{figure*}

The Doppler maps of the representative lines are shown in Figs.~\ref{Fig:dopmaps1}, \ref{Fig:dopmaps2},
and \ref{Fig:dopmapAbs}. Similarly to \citetalias{Paper1}, only out-of-eclipse spectra were used for
tomography.
In addition to different marks on the maps that facilitate interpreting the tomograms (see the captions to the
figures), we also show circles representing velocities at the tidal truncation radius
$r_{\rm max}$, assuming a circular Keplerian flow in the disc. In Section~\ref{Sec:LargeDisc} we discuss
in detail how significant deviations from circular Keplerian velocities can be, taking into account the
gravitational influence of the donor star (see also Fig.~\ref{Fig:Orbits}).
For each line we also show in the middle and right panels the trailed spectra and
their corresponding reconstructed counterparts. To calculate the map of the absorption line \FeII\
5169 (Fig.~\ref{Fig:dopmapAbs}), the latter was first inverted. The complex structure of
the \HeI\ lines requires negative values
in some parts of the Doppler maps to fit the absorption component of the line. To avoid this, we
followed an approach proposed by \citet{MarshUGem} that was also applied by \citet{NeustroevUX}. Prior to the
reconstruction, we have added a Gaussian of FWHM = 600 \kms\ to the data. We then removed the Gaussian
equivalent from the calculated tomogram to produce the final result. This procedure has no effect on the
goodness of the fit or on the entropy when the latter is determined over small scales.

The appearance of the tomograms in superoutburst is quite unusual, although some similarity with the
\Halpha\ Doppler maps in quiescence is apparent (Fig.~\ref{Fig:Dopmap2005}, see also figures 6 and 7 in
\citetalias{Paper1}). No lines show evidence of spiral structures that are sometimes seen in Doppler
maps of CVs in outburst \citep{Steeghs:1997aa}.
Although a diffuse ring of disc emission is still visible, the Balmer and \HeI\ maps are dominated by
a bright emission arc at the right side of the tomograms. In Section~\ref{Sec:Evidence4Ejection} we show
that this arc can probably be regarded as an evolved emission region in the leading side of the accretion
disc, which we discussed in detail in \citetalias{Paper1}. By analogy, hereafter we refer to this structure
as the \emph{\textup{leading arc}} and to its upper right and lower components as the \emph{\textup{upper arc}} and the
\emph{\textup{lower spot}}, respectively. The leading arc follows the circle representing Keplerian velocities at
$r_{\rm max}$. In \Halpha, it is located inside the circle; at its upper part, the arc continues
further to the negative $x$-velocities, where it becomes very bright. This bright, compact, but slightly
elongated emission component (in the following referred to as the \emph{\textup{EEC}}) starts inside the bubble
marking the Roche lobe of the donor star, and moves in the bottom left direction for $\sim$200 \kms.
The EEC \emph{\textup{does not}} follow the predicted trajectory of the gas stream. In \Hbeta, the ECC is much
weaker than in \Halpha,\ and it is not visible in other lines.
In contrast, the \Hbeta\ and especially the \HeI\ lines show another bright spot that is consistent with
the trajectory of the gas stream ($V_{x}$$\approx$$-$900,$V_{y}$$\approx$+100).
A similar spot has clearly been seen in quiescence, and we identified it as the hotspot that is located
well inside the accretion disc (\citetalias{Paper1}). Overall, the emission structure in the Balmer
and \HeI\ lines can be described as having a horseshoe shape with a gap in the third quadrant
($-V_{x}$,$-V_{y}$). Such a gap was also visible in some Doppler maps of HT~Cas in quiescence.
The map of the \SiII\ 6347 line looks different than other lines. In addition to the leading
arc, which is located at the same position as in the \Hbeta\ and \HeI\ maps, there is also another arc
of a larger radius in the third quadrant.

The \HeII\ 4686 emission line is blended with the Bowen blend. This is not a problem for Doppler mapping
because a blend does not generate any compact features on the map. However, it can produce a ring-like
structure \citep{MarshSchwope16}. For the case of HT~Cas, we can simply ignore this effect because the
radius of such a ring on the \HeII\ 4686 map should be $\gtrsim$1700 \kms, which is almost beyond the
edges of the tomogram we showed. \HeII\ 4686 is stronger than \HeI\ 6678 and as strong as \HeI\ 5876, but it
does not show a clear double-peaked profile. As a result, its tomogram exhibits a rather diffuse distribution
of emission with a compact bright spot in the fourth quadrant ($-V_{x}$,$+V_{y}$). The latter can be
identified as the hotspot located in the region where the gas stream hits the outer edge of the accretion
disc. On the other hand, a single-peaked profile suggests non-Keplerian gas motions, for example, perpendicular
to the accretion disc. This indicates that at least part of the \HeII\ 4686 line is formed in the wind
blowing from the disc.

The tomograms of absorption lines (e.g. \FeII\ 5169) show a compact absorption area, which is associated
with the absorption S-wave in the trailed spectra. In the maps it is shifted relative to the position of
the WD in the bottom left direction by a few hundred \kms.
A similar absorption pattern is also seen in most of emission line tomograms;
it might be responsible for the appearance of the gap in the horseshoe emission structure.


\section{Rediscussion of the tidally truncated accretion disc of HT~Cas}
\label{Sec:LargeDisc}

In \citetalias{Paper1} we concluded that the accretion disc radius of HT~Cas in its quiescent
state is very large, close to the tidal truncation radius $r_{\rm max}$.
Our conclusion was based on the measured velocity of the
outer edge of the disc, $V_{\rm out}$, which was accurately measured from the orbit-averaged, double-peaked
profiles of the \Halpha\ emission line. Assuming circular Keplerian rotation, we determined the radius
of the disc to be $R_{\rm d}$=0.52$\pm$0.01$a$, where $a$ is the binary separation. We then compared the
obtained $R_{\rm d}$ with $r_{\rm max}$, which was calculated using Equation 2.61 from \citet{Warner95},
and found that they coincided perfectly.
This result is in contrast to the expectations outlined in the introduction and thus raises the question of how
reliable the assumptions are on which we base our conclusion.
It is known that the gravitational field of the donor star distorts circular particle
orbits (Fig.~\ref{Fig:Orbits}, left-hand panel), thus the assumption of the circular Keplerian flow cannot
be correct in a strict sense \citep{Paczynski}. On the other hand, in \citetalias{Paper1} we discussed the
expected deviation from a circular Keplerian flow and concluded that although the departures at the
outer disc can reach $\sim$20\%\ from Keplerian velocities \citep{Paczynski,SteeghsStehle}, the assumption
of circular Keplerian velocities should still be reliable for the orbit-averaged spectra
because the positive and negative velocity deviations are expected to cancel each other out.

Bearing in mind the importance of the dynamics of the outer disc regions in reaching a conclusion about
the disc radius in HT~Cas, we investigated possible deviations from a circular Keplerian flow in more
detail. For this, we used solutions of the restricted three-body problem because the tidal influence
of the donor on the outer parts of the disc is the primary source of such deviations. We determined
the accurate parameters of a few representative periodic orbits in a binary with the system parameters
of HT~Cas. We note that in the following all the calculated velocities are given for an inclination angle of
81\degr. We first calculated the largest streamline in the restricted three-body problem that does not
intersect any other orbit \citep[we used the code from][]{Ikhsanov04}. It is generally thought that this streamline
represents the limit at which the accretion disc is truncated \citep{Paczynski,IchikawaOsaki94,Warner95}.
In the following we call it the \emph{\textup{truncated}} orbit.\footnote{Despite the simplicity of the
inviscid three-body formalism that was first used by \citet{Paczynski} for calculations of the largest disc size,
other approaches were adopted in a number of studies with very similar results
\citep[see e.g.][]{PapaloizouPringle77,Goodman93,Truss}.} We note that we did not aim to find the
last streamline that starts to cross other orbits. Instead, we reconstructed it for the mass ratio $q$=0.15
using table 1 of \citet{Paczynski}. As expected, the streamline is elongated perpendicular to the line of
centres of the two stars (Fig.~\ref{Fig:Orbits}, left-hand panel). The elongation of orbits rises rapidly when
it approaches the truncation limit. For comparison, we also calculated and show in Fig.~\ref{Fig:Orbits} the
streamlines whose periods $P_{\rm stream}$ are three and four times shorter than the binary orbital period
$P_{\rm orb}$ (3:1 and 4:1 resonance orbits). For the truncated orbit, $P_{\rm orb}$/$P_{\rm stream}$$\approx$2.81.
Following \citet{Paczynski}, the truncated orbit can be characterised by the radii $r_{\rm 1}$=0.434$a$,
$r_{\rm 2}$=0.431$a$, and $r_{\rm max}$=0.522$a$ (Fig.~\ref{Fig:Orbits}, left-hand panel), and the orbit-averaged
radius is $\overline{r}$=0.481$a$. The particle velocity changes along the
streamline, creating an egg-like trajectory in velocity coordinates (Fig.~\ref{Fig:Orbits}, right-hand panel).
Assuming that the accretion disc is truncated by this streamline, this results in periodic modulations of
the peak-to-peak separation of double-peaked emission lines \citep{Paczynski}. The corresponding values of
$V_{\rm out}$ change in the range of 540--670 \kms and have the orbit-averaged value of $\overline{V}_{\rm out}$=605 \kms.
$\overline{V}_{\rm out}$ appears even slightly larger than the \emph{\textup{circular}} Keplerian velocity
$V_{\rm min}$=575 \kms\ at $r_{\rm max}$, the maximum radius-vector of the orbit.\footnote{$r_{\rm max}$
is usually calculated
using Equation 2.61 from \citet{Warner95}. This equation approximates the tabulated values of $r_{\rm max}$
calculated by \citet{Paczynski}. We found that it is accurate to 1\% over the range 0.1 < $q$ < 0.4, but the
accuracy declines significantly outside this limit. While it is not a problem for HT~Cas, we present here
another approximation formula that is accurate to better than 1\% over the range $0.03 < q < 0.73$:
    \begin{equation}
        {r_{\rm max} \over a} = 0.353+0.271\,e^{-3.045\,q}
    \end{equation}
}
Thus, the use of such an easily calculated parameter as $V_{\rm min}$ allows us to place a solid lower limit on
the orbit-averaged $\overline{V}_{\rm out}$ of the tidally truncated disc.
We can conclude now that when the orbit-averaged spectra are discussed, the assumption
of the circular Keplerian flow is still reliable even for very large discs. These calculations strengthen the
main conclusion of \citetalias{Paper1}: the measured $V_{\rm out}^{\rm obs}$=575$\pm$4 \kms\ is lower
than $\overline{V}_{\rm out}$=605 \kms, indicating that the disc edge of HT~Cas in quiescence is at and even
beyond the theoretical truncation limit.

Concluding this section, we note that the question of whether the results from an inviscid calculation
can be applied to real discs was studied in detail by \citet{Truss}. Using hydrodynamic simulations of
viscous accretion discs, \citeauthor{Truss} has shown that the last non-intersecting three-body orbit
remains a good estimate of the tidal truncation radius for $q>0.1$, and hence for HT~Cas.

\section{Evidence for the emitting material beyond the truncation limit}
\label{Sec:Evidence4beyond}

\begin{figure*}
  \centering
  \includegraphics[width=18cm,angle=0]{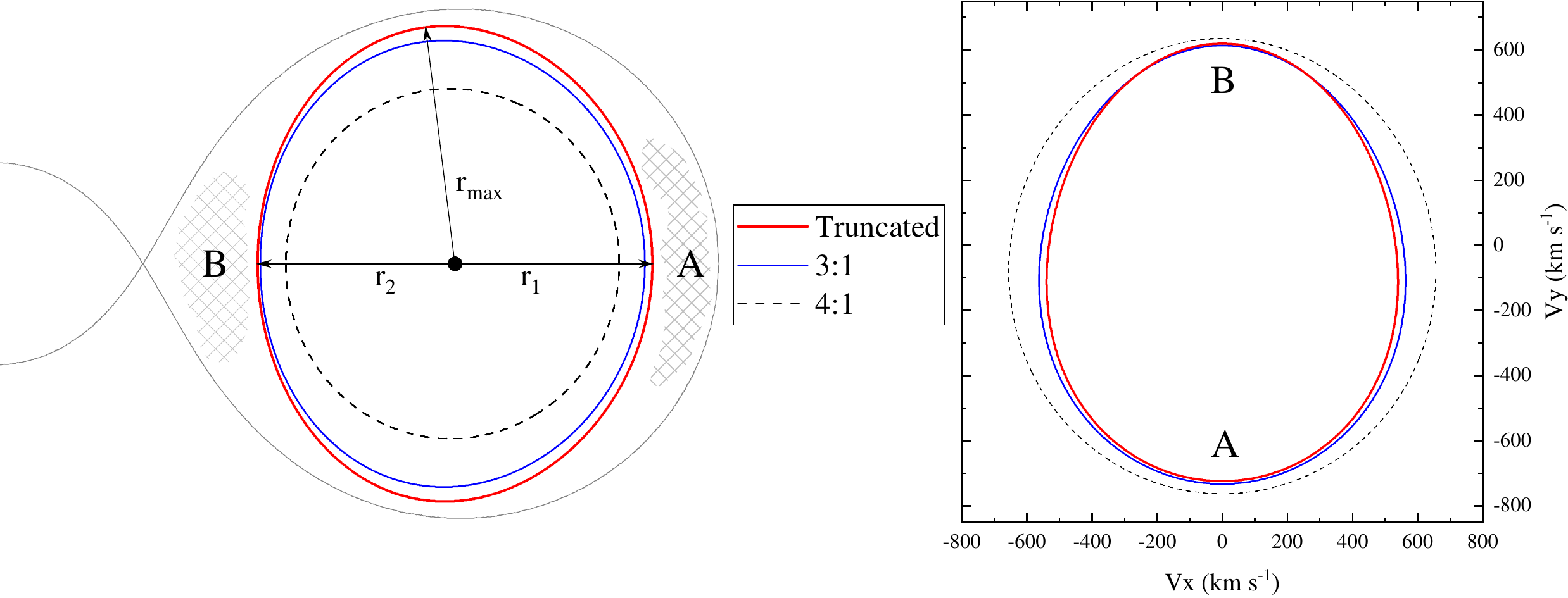}
  \caption{Left: Three representative periodic orbits in the restricted three-body problem calculated for
  a binary with $q$=0.15. The red line shows the largest (truncated)
  streamline that does not intersect any other orbit. The solid blue and dashed black lines show the
  orbits whose periods are three and four times shorter than the binary orbital period (3:1 and 4:1 resonance
  orbits). The Roche lobe is shown with a solid black line.
  $r_{\rm 1}$, $r_{\rm 2}$, and $r_{\rm max}$ are different radii of the streamline.
  Right: Truncated, 3:1, and 4:1 orbits shown in velocity coordinates (Doppler map). A and B
  mark the areas beyond the truncated orbit and their corresponding (roughly) location in the Doppler map.
  See text for detail.
}
  \label{Fig:Orbits}
   \vspace{8mm}
  \centering
   \includegraphics[width=9.93cm,angle=0]{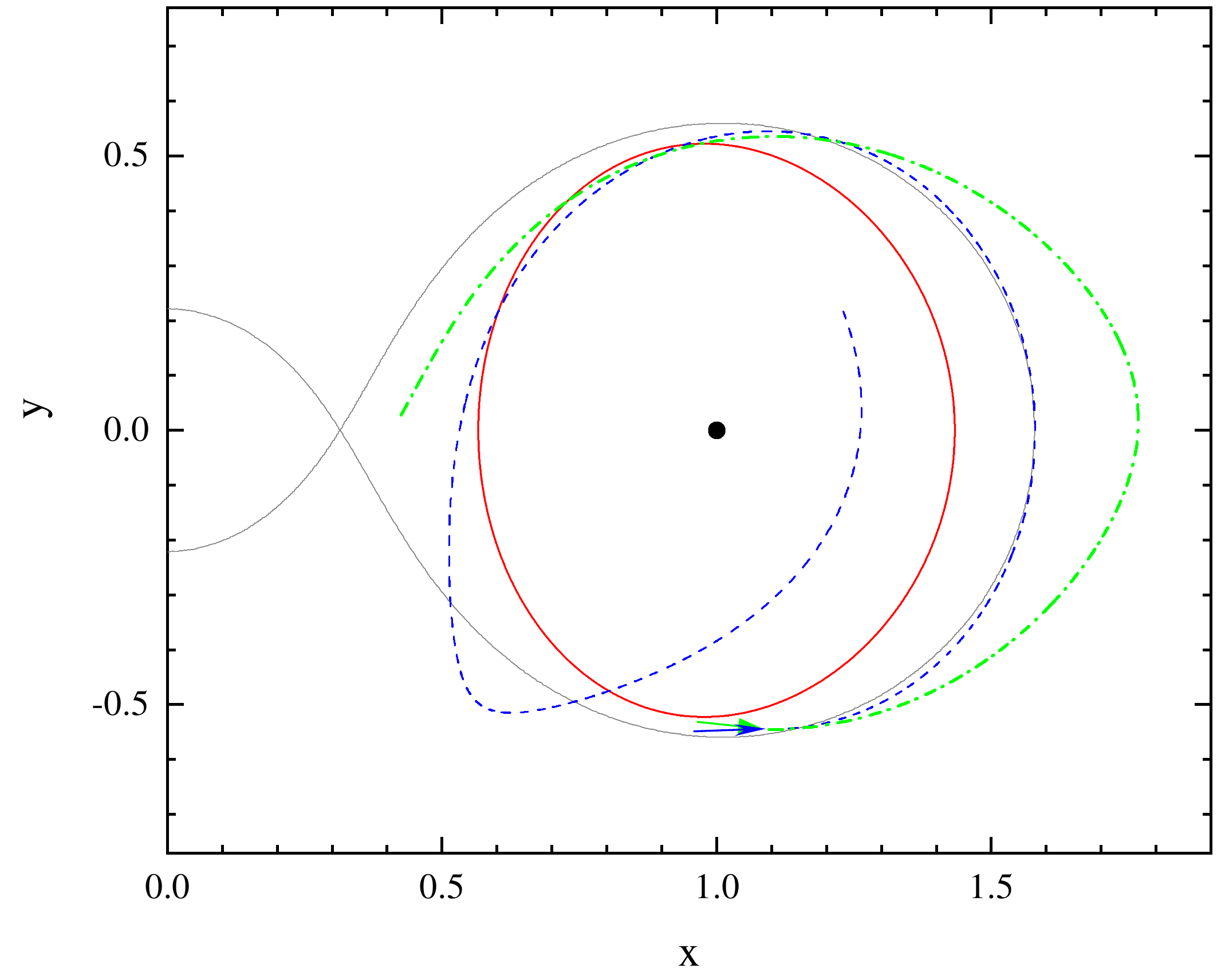}
   \hspace{2mm}
   \includegraphics[width=8.0cm,angle=0]{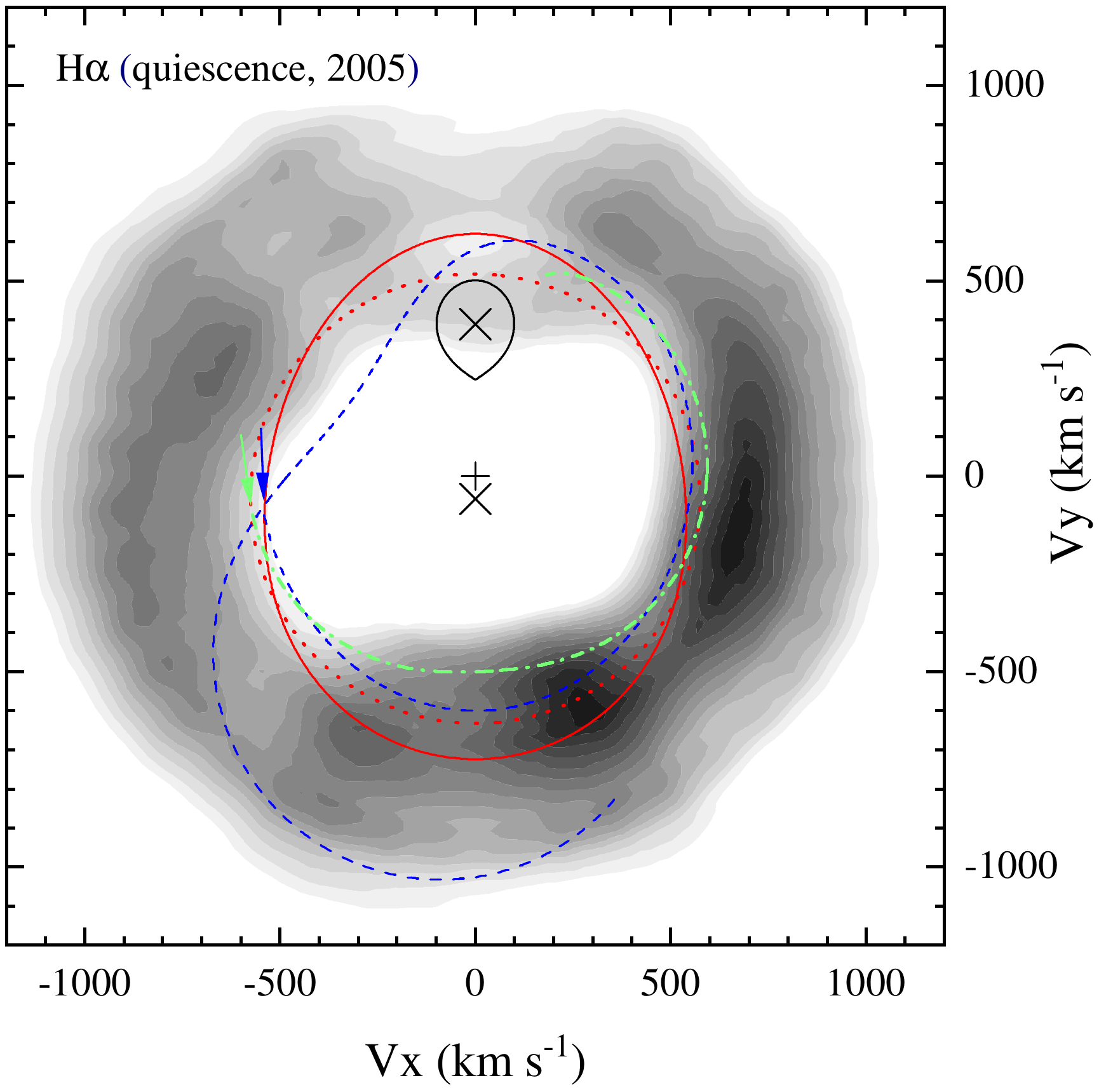}
  \caption{Truncated orbit (solid red line), and two non-periodic trajectories (dashed blue and
            dash-dotted green lines) in the restricted three-body problem calculated for a binary with
            $q$=0.15, shown in spatial (left) and velocity coordinates over the \Halpha\ Doppler map from
            the 2005 observations in quiescence (right). The Doppler map is shown on a linear scale.
            The dotted red line on the map represents circular Keplerian velocities at $r_{\rm max}$.}
  \label{Fig:Dopmap2005}
\end{figure*}

As shown above, the deviation from circular Keplerian motion is quite significant in large discs.
It might be detectable using time-resolved spectroscopy, for instance through Doppler tomography.
In Fig.~\ref{Fig:Dopmap2005} we show the \Halpha\ Doppler map from the 2005 observations of HT~Cas
in quiescence together with the trajectory of the truncated orbit and for comparison with its circular
Keplerian approximation. The emission follows the truncation limit at the left and right
sides of the tomogram, but there is a significant excess of low-velocity emission in the
upper and especially in the bottom parts of the map (e.g. the lower spot is located about 90 \kms\
beyond the truncation limit).

The exact spatial positions of these areas is difficult to establish. They roughly correspond to the
areas A and B in Fig.~\ref{Fig:Orbits} in which no stable streamlines are possible and so no accurate
inverse transformation from velocity to spatial coordinates. However, it can be shown that quasi-circular
orbits that extend beyond the truncation limit are consistent with the observed excess of low-velocity
emission (Fig.~\ref{Fig:Dopmap2005}, right-hand panel).

The appearance of material outside the truncation limit can have different reasons. In this
respect, we must admit that the real shape of accretion discs should be more complex than that of the
truncated orbit. First of all, real discs are not inviscid, and viscosity plays a role in the transport
of angular momentum and hence in the disc truncation; some deviations from the three-body
calculations are not excluded. It was shown that at some physical conditions, the disc can extend beyond
the truncated orbit, although only slightly \citep[see e.g.][]{PapaloizouPringle77}.
Moreover, the truncated orbit in HT~Cas is larger than (although very close to) the 3:1 resonance orbit
(Fig.~\ref{Fig:Orbits}, left-hand panel)\footnote{The
3:1 resonance orbit is characterized by the following parameters: $r_{\rm 1}$=0.417, $r_{\rm 2}$=0.428,
$r_{\rm max}$=0.492, the orbit averaged radius is $\overline{r}$=0.459. $\overline{r}$ has the same value
as that calculated in the usual manner using the formula $r_{\rm 3:1} = a\,(1/3)^{2/3}\,(1+q)^{-1/3}$.
The corresponded values of $V_{\rm out}$
change in the range of 560--670 \kms, having the orbit-average value of $\overline{V}_{\rm out}$=617 \kms.},
which is believed to be tidally unstable \citep{WhitehurstKing91}. Smoothed particle hydrodynamics (SPH)
models predict that at the 3:1 resonance radius the disc becomes eccentric and precessing \citep{Smith07}.
The SPH calculations show that such a disc has quite a complex shape, and sometimes, ejected streams of
matter fill the gap between the tidal orbit and the L1 and L3 Lagrangian points.
On the other hand, it has been shown that the tidal instability does not occur in Eulerian models, which
use a full energy equation instead of an isothermal approximation \citep[][Bisikalo, priv. comm.]{KornetRozyczka}.
It is interesting that the latter models often show the appearance of disc structures consistent with
the lower spot in HT~Cas \citep[see e.g. figure 6 in][]{KornetRozyczka}.

It is important to point out that the \HeI\ and higher order Balmer lines in outburst follow the predicted
truncated orbit very accurately (see the following section). This confirms that the truncated limit adopted
in this paper is well justified, supporting the conclusion that there is the emitting material in the quiescent
disc of HT~Cas beyond the truncation limit.

\begin{figure*}
\centerline{
  \includegraphics[height=8cm]{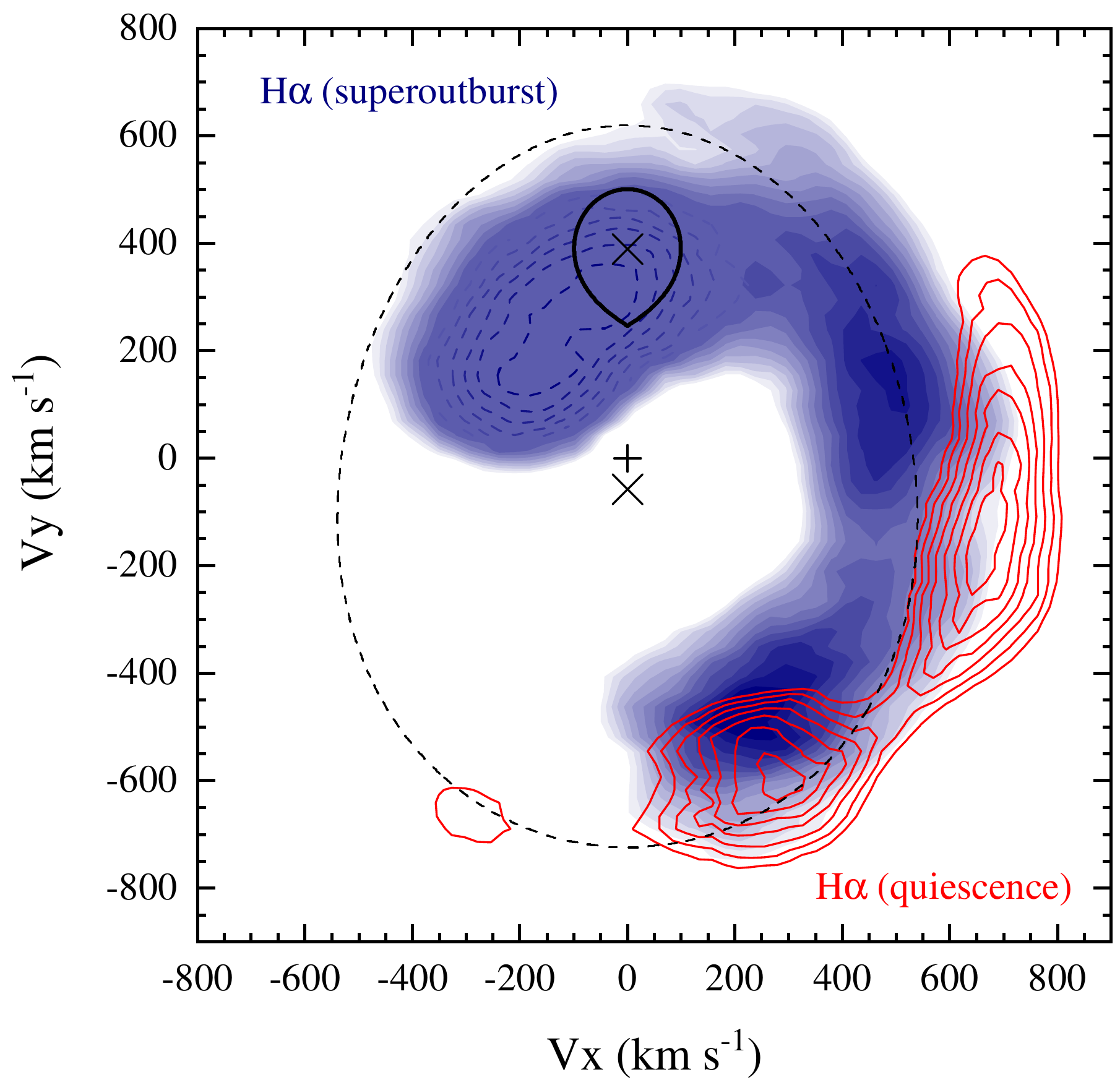}
  \includegraphics[height=8cm]{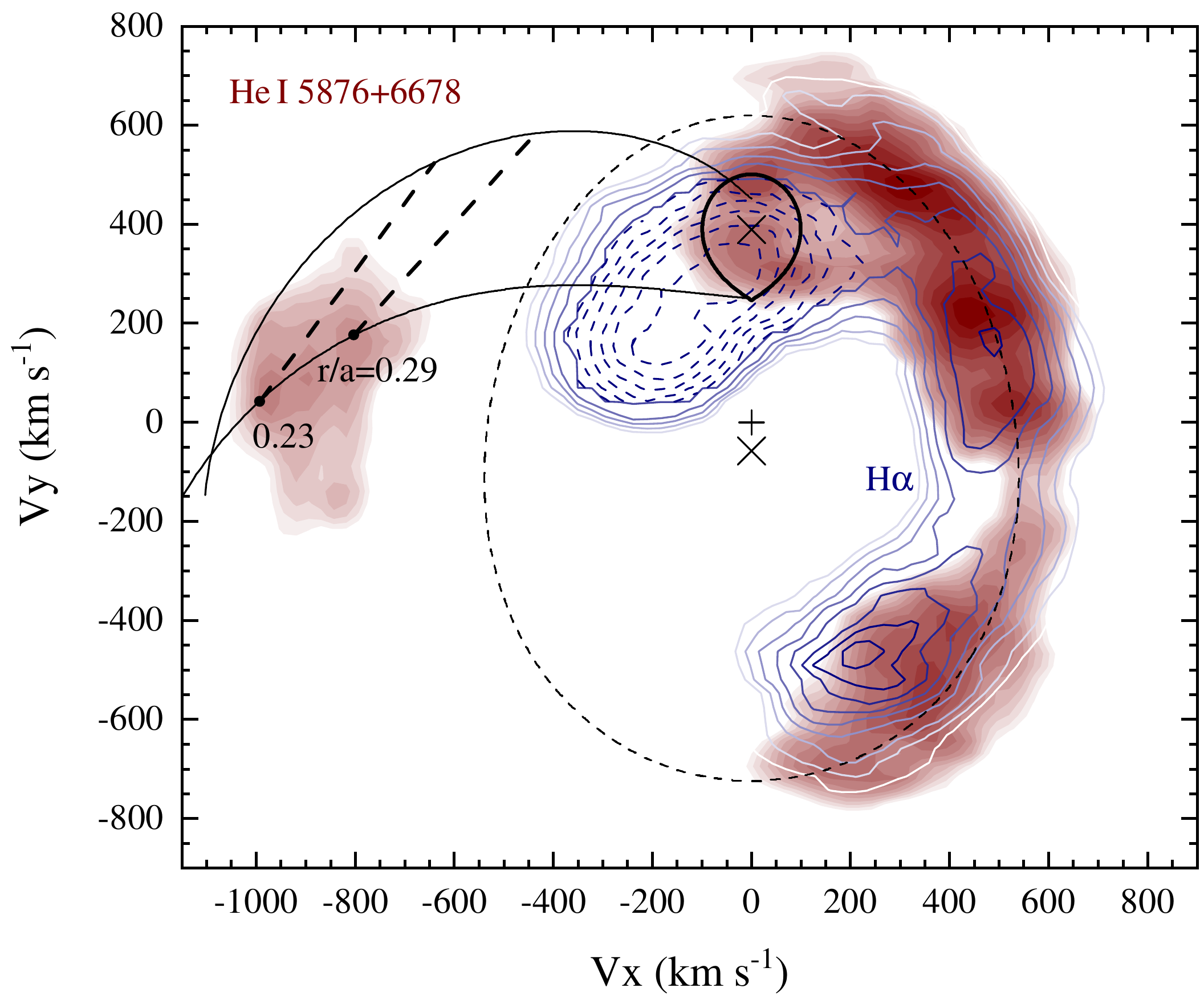}
}
\caption{Doppler maps combined from the tomograms for the \Halpha\ line in superoutburst and quiescence
         (left) and for \Halpha\ and the \HeI\ lines in superoutburst (right).
         The maps are shown on a linear scale. Different colours represent different lines (red and
         navy show \Halpha\ in quiescence and in superoutburst, respectively, and dark red shows the \HeI\ lines
         in superoutburst). The EEC, the brightest emission component in \Halpha\ in superoutburst,
         is flattened to show other emission structures in more detail. The quasi-elliptical dashed
         lines represent velocities at the tidal truncation limit.
         }
\label{Fig:ZoomedMaps}
\end{figure*}

\section{Evidence for the emitting material outside the WD Roche lobe during the superoutburst}
\label{Sec:Evidence4Ejection}

The common structure that is clearly seen in the outburst tomograms resembles the most prominent emission
area of HT~Cas in quiescence: the leading arc.
In quiescence, however, this area is most obvious
in the \Halpha\ line alone, whereas in \Hbeta\ it is much weaker, and it is almost undetectable in the \HeI\ lines.
In superoutburst, in contrast, it is the brightest region in both the \Hbeta\ and \HeI\ lines, and it is bright
in \SiII\ 6347. In Section~\ref{Sec:Evidence4beyond} we discussed the spatial location of the leading arc
components in quiescence. It is instructive to examine how their
location has changed in superoutburst.

Figure~\ref{Fig:ZoomedMaps} compares the location of the leading arc that is visible in \Halpha\
in superoutburst and in quiescence (left-hand panel) to that found in \HeI\ lines in superoutburst (right-hand
panel). First of all, we point out that both components of this emission region in the \HeI\ lines (this also
holds for \Hbeta\ and \SiII, see Fig.~\ref{Fig:dopmaps1}) closely follow the truncation limit, with
some extension beyond it. The fact that different emission lines that cover a range of excitation energies and
temperatures accurately trace the predicted truncation limit strongly indicates that the accretion disc of
HT~Cas is indeed truncated and that it is relatively hot ($\sim$15000~K) until the very edge in superoutbursts.

In this respect, it is extremely interesting that during the superoutburst, both the emission components
of \Halpha\ were shifted toward lower velocities and thus appeared well inside the truncation limit
in velocity space
(Fig.~\ref{Fig:ZoomedMaps}, left-hand panel). The deviation of the upper arc from the limit is at least
100 \kms, and of the lower spot, it is at least 200 \kms. The problem is that the upper arc in the \Hbeta\ and
\HeI\ lines is situated in the most elongated area of the truncated orbit, which almost touches the Roche-lobe
surface (Fig.~\ref{Fig:Orbits})\footnote{In HT~Cas, the equatorial Roche-lobe radius (in the direction
perpendicular to the line of centres of the WD and the donor) is 0.558$a$, and $r_{\rm max}$=0.522$a$.}.
It leaves no room for low-velocity \Halpha\ emission, which could only originate \emph{\textup{outside}} the WD Roche lobe.
Assuming a Keplerian flow around the centre of mass of a binary, the distance would be at least 1.5$\times
r_{\rm max}$. We point out, however, that no single orbit in the restricted three-body problem can fully
explain the observed emission structures all the way from the lower spot to the upper part of the upper arc.
Thus, this is only a rough estimate.

In order to place the \Halpha\ emission inside the WD Roche lobe, we need to assume significantly
different velocities than predicted in the restricted three-body problem, and that \Halpha\ \emph{\textup{does not follow}}
the dynamics that \Hbeta\ and other spectral lines \emph{\textup{follow}}. Taking into account the previous discussion
and the work by \citet{Truss}, this appears highly unlikely.
The appearance of the \Halpha\ emission in the lower velocity region, in contrast with other, hotter lines,
can be interpreted as tracing a cooling gas ($\lesssim$10\,000~K) that is expelled from the hot accretion disc
through a sector shown in green in Fig.~\ref{Fig:HTCas}.

\section{Other emission and absorption components}

\subsection{Elongated emission component }

The cool gas region visible in \Halpha\ envelopes the binary for at least
180\degr\ in azimuth and visually curves in at the top part of the tomogram to even lower velocities.
Nevertheless, it is not clear whether the EEC at the top of the Doppler map and the leading arc are
physically linked. One side of the EEC coincides in the Doppler map with the
Roche lobe of the donor star. This allows us to speculate about the possible origin of this part of the EEC
in the inner semi-sphere of the donor star, which may have been irradiated by the WD and/or hot
accretion disc regions.
However, although the EEC is the brightest emission area seen in \Halpha, its other part (a trajectory)
cannot be associated with any structures of the binary system such as the gas stream or the hotspot, and
none of the hydrodynamical simulations predict an increase in emission in this region. On the other hand,
if the EEC is not locked in the rotating binary frame, then its low velocities suggest an origin
outside the Roche lobe of the WD. In this respect, we note that if the expelled material leaves the
system, then it will create a circumbinary envelope \citep[this resembles the propeller mechanism used to explain
Doppler maps of AE~Aqr, see figures 1--3 in][]{Ikhsanov04}, the velocity of which are roughly consistent with
the EEC. However, the distribution of the material surrounding the system is expected to have an azimuthal
symmetry, and we have no physical explanation why we observe only the EEC instead of the ring, and why it
is so bright.

\subsection{Non-axisymmetrical absorption}
\label{Sec:Absorption}

The relative compactness of the absorption structure in the Doppler maps suggests that its source is locked
in the binary rest frame. Assuming Keplerian velocities, it is located opposite of the donor
star, at side of the disc at azimuth $\sim$110$\degr$. To be eclipsed by the donor, the absorption
region must be situated close to the WD and it should not be very extended vertically
(see Fig.~\ref{Fig:HTCas}, left-hand panel). This resembles the so-called dark spot phenomenon that
has been observed in the
nova-like UX~UMa, which has been explained in the context of the stream-disc overflow model
\citep[see][and references therein]{NeustroevUX}.
However, this model cannot explain the depression of the blue peak of the emission lines at orbital phases
$\sim$0.3$\pm$0.1, when the absorption component is the most blueshifted (Fig.~\ref{Fig:Absorptions},
upper right panel). These two phenomena are probably not physically related.

Such a depression might occur if the very edge of the accretion disc, perpendicular to the line of sight,
is temporarily obscured by something in the foreground, for example by a thickened sector of the outer
disc (Fig.~\ref{Fig:HTCas}, right-hand panel). This model is consistent with our hypothesis that
the bright region on the leading side of the disc in quiescence (and the leading arc in outburst) is
caused by irradiation of
tidally thickened sectors of the outer disc by the WD and/or hot inner disc regions (see
\citetalias{Paper1} and figure 10 therein). See Section~\ref{Sec:Discussion} for further discussion.

\subsection{Hotspot}

The canonical hotspot from the area of interaction between the gas stream and the disc is often one of
the brightest emission components of dwarf novae in quiescence. This is, however, rarely visible during
outbursts, when the hotspot cannot reach a high contrast that would allow it to be detected on the background of the hot and
luminous disc.
Surprisingly, the hotspot in HT~Cas in outburst is easily detected in most of the emission lines (the only
exception is \Halpha). However, only in \HeII\ is the spot located at the very edge of the disc, whereas
in other lines, it is situated well inside the disc, at distances from the WD of
$R_{\rm hs}$$\approx$0.23--0.29$a$. These distances are consistent with those found for the hotspot that is visible
in different lines during the quiescent state (\citetalias{Paper1}). The difference is that in quiescence,
the hotspot velocities were close to \emph{Kepler} velocities at the spot position \citepalias[see the left-hand
panel of fig. 11 in][]{Paper1}, while during the superoutburst, the velocities were close to the expected
velocity of the gas stream at $R_{\rm hs}$ (Figs.~\ref{Fig:dopmaps1} and \ref{Fig:dopmaps2}, and the
right-hand panel of Fig.~\ref{Fig:ZoomedMaps}).

Following \citet{WZ1} and \citet{WZ2}, we explained in \citetalias{Paper1} the appearance of the
hotspot inside the disc by a very low density of the outer disc regions that allows the gas stream
to penetrate deep into the disc. The occurrence of the hotspot at the same location during the superoutburst
points to the same explanation, but it also suggests relatively low temperatures in the region of the
hotspot origin. The latter is not expected for the disc in the middle of the superoutburst when it is in
a quasi-stable high state.

\section{Discussion}
\label{Sec:Discussion}

It is commonly accepted that variations in the outer disc radius in interacting binaries play an important
role in understanding the structure and evolution of accretion discs. These variations are predicted by
various models of discs \citep[see e.g.][]{Smak84,Lasota2001,HameuryLasota05}, and they are adopted to
explain superhumps that are observed during superoutbursts in SU~UMa-type dwarf novae \citep{Osaki05}. Here
we mention two important theoretical results obtained by \citet{HameuryLasota05}. They found significant
variations in the outer disc radius during an outburst (at least 20\%), and that the tidal torques that
determine the outer radius at which the disc is truncated must be important also well inside $r_{\rm max}$.
Changes in eclipse widths and eclipse contact times for the bright spot that were detected for several
dwarf novae are usually given as observational evidence of disc radius variations \citep{Smak96}. For instance,
this approach allowed \citet{Patterson81} to deduce that the disc of HT~Cas expands by $\sim$50\% during outbursts.

In contrast to these claims, we have shown in \citetalias{Paper1} that the disc radius of HT~Cas in quiescence
was nearly the same during many years of observations. Moreover, we here demonstrated that
during the superoutburst, the disc of HT~Cas has become hot but remained the \emph{\textup{same}} size. This conclusion
is independent of the velocity field in the accretion disc and only assumes that this field is globally
stable over time. The important empirical fact is that assuming Keplerian rotation,
this quasi-constant radius of the disc appears to be close to the tidal truncation limit.
This limit in short-period CVs is always larger
than the 3:1 resonance radius. This result questions the standard explanation that the
appearance of superhumps is a result of the disc radius expansion beyond the 3:1 resonance radius
\citep{Osaki96}.

Admitting that the assumption of the circular Keplerian rotation may not be fully correct,
we calculated the last non-intersecting three-body orbit,
which was shown to be a good estimate of the tidal truncation limit for the system parameters of HT~Cas
\citep{Truss}. One of our conclusions from these calculations is that the assumption of the circular Keplerian
flow is still reliable when the orbit-averaged spectra are analysed. It is even more appealing that when the
truncated orbit is visualised in a Doppler map, it traces the leading emission arc even more accurately than
for the case of circular orbits. This further confirms our earlier suggestion
that the disc of HT~Cas is truncated at the tidal truncation limit in both quiescence and outburst.

How does this result depend on the assumption that gravitational forces prevail in the disc?
Obviously, if the basic assumption of orderly, gravitational motion breaks down, the emission may not arise
where it does according to solutions of the restricted three-body problem. However, we argue that this will
not affect our main conclusion that the disc in HT~Cas, in both quiescence and outburst, remains near
the theoretical maximum radius given by the truncation limit. Observations in quiescence show that
the \HeI\ and higher order Balmer lines are broader than \Halpha, \emph{\textup{differently}} for each line. This is
explained by the fact that they originate in higher temperature areas of the cool disc, closer to the WD.
In outburst, however, the width and peak-to-peak separation of most of
the lines became very similar to each other and close to those of \Halpha\ in quiescence. They also share the
\emph{\textup{same}} position in the Doppler maps as \Halpha\ in quiescence, meaning that the disc is indeed
truncated here.
Moreover, the fact that different emission lines that cover a range of excitation energies and
temperatures accurately trace the predicted truncation limit (Figs.~\ref{Fig:dopmaps1}-\ref{Fig:dopmaps2} and
\ref{Fig:ZoomedMaps})
can be interpreted as a sign that the deviations from gravitationally dominated motion of the accretion flows
are not very significant\footnote{We also refer to the recent work by \citet{DoganNixon}.
They analysed effects of non-Keplerian rotation
in accretion discs and concluded that the departure from Keplerian rotation in even the outer disc regions in
CVs is quite small.}, although we cannot completely exclude them.

The case of HT~Cas is not unique. In \citetalias{Paper1} we inspected
several CVs with relatively well measured system parameters and found that the accretion disc in most of them
(\object{VW~Hyi}, \object{WZ~Sge}, \object{V406~Vir}, \object{EZ~Lyn}, \object{RR~Pic}, \object{and IP~Peg}) also
has a radius that is close to the tidal truncation limit. Moreover, it has been reported that in WZ~Sge the same disc
radius has been observed for 40 years \citep{WZ2}. All this strengthens our hypothesis that the accretion disc
in CVs and possibly in other interacting binaries is \emph{\textup{always}} extended to its truncation
limit. Much evidence exists that the outermost regions of the disc have quite a low density and much lower optical
depth than the inner disc, and they contribute only a tiny fraction to the total optical flux of the system.
The ``observed'' radius variations are thus the variations of the ``photometric'' disc edge, which can move
depending on the physical conditions in the disc. However, beyond this edge, there is still material that
repeatedly orbits the WD and mostly contributes to emission lines rather than to broadband continuum light
(see \citetalias{Paper1} for a further discussion).

We have also presented several lines of evidence that indicate that the \Halpha\ emission during the
superoutburst most probably originated beyond the Roche lobe of the WD. We propose that we observe here
relatively cool material that was ejected from the leading side of the hot disc during the superoutburst.
\Halpha\ traces cooler gas with lower ionisation than other Balmer and helium lines; this gas in the outflow
region can be easily excited by the hot accretion disc to produce the \Halpha\ emission line. The amount
of expelled matter must be quite significant to be able to obscure the disc edge perpendicular to the
line of sight (Fig.~\ref{Fig:HTCas}, right-hand panel). This hypothesis is consistent with predictions
of numerical simulations, in which the appearance of the circumbinary envelope created by matter that left
the accretion disc and went outside the Roche lobe has been noted \citep[see e.g.][]{Bisikalo98}.
An expansion of material outside the Roche lobe of the accretor was predicted even for systems in the
quiescent state. Synthetic Doppler maps of such simulations \citep{Bisikalo08} closely resemble the
observed tomograms of HT~Cas.

This result can have important implication for CV evolution theory, whose
predictions are not fully supported by observations of the currently known CV sample.\footnote{Here we
note the following discrepancies between the available models of CV evolution and the observations
relevant to this paper:
1) the measured effective temperatures of WDs in short-period CVs show a large scatter and imply higher
mass-transfer rates than predicted \citep{Pala17}; 2) although the current empirical period distribution
reveals a significant accumulation of systems near the observed minimum period \citep{Gaensicke09,Kato17}
that resembles the period-minimum spike predicted by CV population models \citep{KolbBaraffe99}, the
width of the empirical spike is larger and the relative number of CVs at the period minimum is much smaller
than predicted \citep{Knigge11,Pala19}.} The
evolution of a CV is driven by angular momentum loss from the binary. According to the standard model,
angular momentum loss in short-period CVs (below the period gap) is assumed to be driven solely by
gravitational radiation \citep[for details, see][]{Knigge11}. However, the
material that leaves the system carries with it the specific orbital angular momentum of the WD. This
material is expected to flow out within the orbital plane of the system and thus can create and feed
(continuously or occasionally, e.g. during outbursts) a circumbinary disc (CB). It has been shown that
the inclusion of angular momentum loss associated with a CB disc can significantly affect the evolution
of CVs \citep{SpruitTaam01,TaamSpruit01}. In particular, \citet{TaamSandquistDubus03} have shown that
even low fractional mass input rates into the CB ($\delta$$\sim$10$^{-4}$) can promote the mass transfer
between the binary components and thus increase the evolution rate and heat the WD by the increased
infall of material. Moreover, assuming even lower $\delta$$\sim$10$^{-5}$, \citet{Willems05} have
demonstrated that this form of angular momentum loss tends to smooth the predicted spike in the
number of systems near the period minimum. Deeper discussion of the possible effect of a CB on the CV
evolution is beyond the scope of this paper.

How significant and how common is mass outflow from the outer accretion disc in CVs and related objects?
Some evidence for a possible expansion of accretion disc material beyond the Roche lobe has been presented
for long-period nova-like CVs \citep[see e.g.][]{RWSex,RWTri}. However, to the best
of our knowledge, the observational confirmation of such an expansion in a short-period CV is obtained
for the first time. It was possible because HT~Cas is an eclipsing binary and its system parameters
are known accurately enough, and because it exhibited emission lines during the superoutburst, which
is not a common case. Unfortunately, only a few short-period CVs were observed spectroscopically during
their superoutbursts, and most of them are non-eclipsed.
However, in a few WZ~Sge-type dwarf nova exhibiting double-peaked emission lines in superoutburst, the
peak-to-peak separation of \Halpha\ was found to be significantly smaller than that of other lines
(V1838 Aql,  \citealt{V1838Aql}, and TCP J21040470+4631129, \citealt{Teyssier19,NeustroevTCP1}).
Moreover, in TCP J2104070+4631129, all the Balmer and \HeI\ lines are much broader in quiescence
than they were during the superoutbursts \citep{NeustroevTCP2}.
We also noted the presence of a horseshoe structure in the \Halpha\ Doppler map of SSS J122221.7$-$311525
\citep[][Neustroev et al., in prep]{NeustroevSSS}. It is very interesting that this system just after
the end of the superoutburst plateau showed the appearance of a strong near-infrared excess resulting in very red
colours. The colours then became bluer again, but it took a few hundred days to acquire a stable level.
Similar reddening of optical light after the superoutburst has been reported for several other WZ Sge-type
stars \citep[see][and references therein]{NeustroevSSS}.
All these features probably indicate mass outflow from the outer disc.
Unfortunately, the presented observations do not allow for a reasonable estimate of the mass-loss rate
through the outer disc. An attempt to estimate this parameter might be made by applying
numerical simulations.

\section{Summary}

We have analysed time-resolved spectroscopic observations of the dwarf nova HT~Cas during its 2017
superoutburst with the aim of comparing the properties of the accretion disc in the system during
superoutburst and in quiescence. We also discussed again the location of emission structures and the
accretion disc size in quiescence using solutions of the restricted three-body problem.
The principal results of this study are summarised below.

\begin{enumerate}
    \item The superoutburst spectrum is similar in appearance to the quiescent spectra, although
    the strength of most of the double-peaked emission lines of the Balmer series and \HeI\ decreased.
    In addition, the high-excitation lines of \HeII\ 4686, \CII\ 4267, and the Bowen blend significantly
    strengthened in comparison with the Balmer lines.
    We also detected the emission lines of \SiII,\ which are rarely observed in CVs.

    \item Many lines show a mixture of broad emission and narrow absorption components.
    In the \HeI\ lines, the absorption extends below the continuum. The iron lines are dominated by
    the absorption, or their emission component is very weak. In the \Halpha\ and \Hbeta\ lines,
    the absorption is undetectable in the mean spectrum, although its sign is visible in the trailed
    spectra.

    \item \Halpha\ in outburst was much narrower than in quiescence. Other emission lines also narrowed
    in outburst, but they did not become as narrow as \Halpha.

    \item The single-peaked profile of \HeII\ suggests that at least part of this line is formed in the
    wind blowing from the disc.

    \item Doppler maps of the Balmer and \HeI\ lines
    are dominated by a bright emission arc at the right side of the tomograms,
    superposed on a diffuse ring of emission. This leading arc is probably an evolved emission region
    at the leading side of the accretion disc; this is the dominant emission source of the \Halpha\ line
    in quiescence.

    \item In \Halpha\ in quiescence, and in the \Hbeta, \HeI, and \SiII\ lines in superoutburst,
    the leading arc is located at and even beyond the theoretical truncation limit, indicating thus that
    the disc size during the superoutburst remained the same as in quiescence.

    \item Doppler tomography of \Halpha\ in superoutburst revealed that the bulk of its emission, which
    traces cooler gas than other studied lines, is produced beyond the Roche lobe of the WD. We interpret this as a
    signature that cooling gas is expelled from the hot disc.
\end{enumerate}

These unexpected findings question the standard explanation for superoutbursts and the
appearance of superhumps. They can have important implications for CV evolution theory.

\begin{acknowledgements}
We are thankful to the anonymous referees whose comments helped greatly to improve the paper.
We acknowledge the financial support from the visitor and mobility program of the Finnish Centre for
Astronomy with ESO (FINCA), funded by the Academy of Finland grant No. 306531. This work was supported
by PAPIIT grant IN-102120. Based on observations made with the Nordic Optical Telescope, operated by
the Nordic Optical Telescope Scientific Association at the Observatorio del Roque de los Muchachos,
La Palma, Spain, of the Instituto de Astrofisica de Canarias. The data presented here were obtained
with ALFOSC, which is provided by the Instituto de Astrofisica de Andalucia (IAA) under a joint
agreement with the University of Copenhagen and NOTSA.
We acknowledge with thanks the variable star observations from the AAVSO International Database
contributed by observers worldwide and used in this research.
\end{acknowledgements}

\bibliographystyle{aa} 


\begin{thebibliography}{59}
\expandafter\ifx\csname natexlab\endcsname\relax\def\natexlab#1{#1}\fi

\bibitem[{{Bisikalo} {et~al.}(1998){Bisikalo}, {Boyarchuk}, {Chechetkin},
  {Kuznetsov}, \& {Molteni}}]{Bisikalo98}
{Bisikalo}, D.~V., {Boyarchuk}, A.~A., {Chechetkin}, V.~M., {Kuznetsov}, O.~A.,
  \& {Molteni}, D. 1998, \mnras, 300, 39

\bibitem[{{Bisikalo} {et~al.}(2004){Bisikalo}, {Boyarchuk}, {Kaigorodov},
  {Kuznetsov}, \& {Matsuda}}]{Bisikalo04}
{Bisikalo}, D.~V., {Boyarchuk}, A.~A., {Kaigorodov}, P.~V., {Kuznetsov}, O.~A.,
  \& {Matsuda}, T. 2004, Astronomy Reports, 48, 588

\bibitem[{{Bisikalo} {et~al.}(2008){Bisikalo}, {Kononov}, {Kaigorodov},
  {Zhilkin}, \& {Boyarchuk}}]{Bisikalo08}
{Bisikalo}, D.~V., {Kononov}, D.~A., {Kaigorodov}, P.~V., {Zhilkin}, A.~G., \&
  {Boyarchuk}, A.~A. 2008, Astronomy Reports, 52, 318

\bibitem[{{Borges} {et~al.}(2008){Borges}, {Baptista}, {Papadimitriou}, \&
  {Giannakis}}]{Borges2008}
{Borges}, B.~W., {Baptista}, R., {Papadimitriou}, C., \& {Giannakis}, O. 2008,
  \aap, 480, 481

\bibitem[{{Do{\u{g}}an} \& {Nixon}(2020)}]{DoganNixon}
{Do{\u{g}}an}, S. \& {Nixon}, C.~J. 2020, \mnras, 495, 1148

\bibitem[{{Feline} {et~al.}(2005){Feline}, {Dhillon}, {Marsh}, {Watson}, \&
  {Littlefair}}]{Ultracam}
{Feline}, W.~J., {Dhillon}, V.~S., {Marsh}, T.~R., {Watson}, C.~A., \&
  {Littlefair}, S.~P. 2005, \mnras, 364, 1158

\bibitem[{{G{\"a}nsicke} {et~al.}(2009){G{\"a}nsicke}, {Dillon}, {Southworth},
  {Thorstensen}, {Rodr{\'{\i}}guez-Gil}, {Aungwerojwit}, {Marsh}, {Szkody},
  {Barros}, {Casares}, {de Martino}, {Groot}, {Hakala}, {Kolb}, {Littlefair},
  {Mart{\'{\i}}nez-Pais}, {Nelemans}, \& {Schreiber}}]{Gaensicke09}
{G{\"a}nsicke}, B.~T., {Dillon}, M., {Southworth}, J., {et~al.} 2009, \mnras,
  397, 2170

\bibitem[{{Goodman}(1993)}]{Goodman93}
{Goodman}, J. 1993, \apj, 406, 596

\bibitem[{{Hameury}(2019)}]{Hameury19}
{Hameury}, J.-M. 2019, arXiv e-prints, arXiv:1910.01852

\bibitem[{{Hameury} \& {Lasota}(2005)}]{HameuryLasota05}
{Hameury}, J.~M. \& {Lasota}, J.~P. 2005, \aap, 443, 283

\bibitem[{{Hernandez} {et~al.}(2017){Hernandez}, {Zharikov}, {Neustroev}, \&
  {Tovmassian}}]{RWSex}
{Hernandez}, M.~S., {Zharikov}, S., {Neustroev}, V., \& {Tovmassian}, G. 2017,
  \mnras, 470, 1960

\bibitem[{{Hern{\'a}ndez Santisteban} {et~al.}(2019){Hern{\'a}ndez
  Santisteban}, {Echevarr{\'{\i}}a}, {Zharikov}, {Neustroev}, {Tovmassian},
  {Chavushyan}, {Napiwotzki}, {Costero}, {Michel}, {S{\'a}nchez},
  {Ruelas-Mayorga}, {Olgu{\'{\i}}n}, {Garc{\'{\i}}a-D{\'{\i}}az},
  {Gonz{\'a}lez-Buitrago}, {de Miguel}, {de la Fuente}, {de Anda}, \&
  {Suleimanov}}]{V1838Aql}
{Hern{\'a}ndez Santisteban}, J.~V., {Echevarr{\'{\i}}a}, J., {Zharikov}, S.,
  {et~al.} 2019, \mnras, 486, 2631

\bibitem[{{Horne} \& {Marsh}(1986)}]{HorneMarsh86}
{Horne}, K. \& {Marsh}, T.~R. 1986, \mnras, 218, 761

\bibitem[{{Horne} {et~al.}(1991){Horne}, {Wood}, \& {Stiening}}]{Horne91}
{Horne}, K., {Wood}, J.~H., \& {Stiening}, R.~F. 1991, \apj, 378, 271

\bibitem[{{Ichikawa} \& {Osaki}(1994)}]{IchikawaOsaki94}
{Ichikawa}, S. \& {Osaki}, Y. 1994, \pasj, 46, 621

\bibitem[{{Ikhsanov} {et~al.}(2004){Ikhsanov}, {Neustroev}, \&
  {Beskrovnaya}}]{Ikhsanov04}
{Ikhsanov}, N.~R., {Neustroev}, V.~V., \& {Beskrovnaya}, N.~G. 2004, \aap, 421,
  1131

\bibitem[{{Kafka}(2017)}]{AAVSO}
{Kafka}, S. 2017, Observations from the AAVSO International Database,
  https://www.aavso.org

\bibitem[{{Kato} {et~al.}(2017){Kato}, {Isogai}, {Hambsch}, {Vanmunster},
  {Itoh}, {Monard}, {Tordai}, {Kimura}, {Wakamatsu}, {Kiyota}, {Miller},
  {Starr}, {Kasai}, {Shugarov}, {Chochol}, {Katysheva}, {Zaostrojnykh},
  {Seker{\'a}{\v{s}}}, {Kuznyetsova}, {Kalinicheva}, {Golysheva}, {Krushevska},
  {Maeda}, {Dubovsky}, {Kudzej}, {Pavlenko}, {Antonyuk}, {Pit}, {Sosnovskij},
  {Antonyuk}, {Baklanov}, {Pickard}, {Kojiguchi}, {Sugiura}, {Tei}, {Yamamura},
  {Matsumoto}, {Ruiz}, {Stone}, {Cook}, {de Miguel}, {Akazawa}, {Goff},
  {Morelle}, {Kafka}, {Littlefield}, {Bolt}, {Dubois}, {Brincat}, {Maehara},
  {Sakanoi}, {Kagitani}, {Imada}, {Voloshina}, {Andreev}, {Sabo}, {Richmond},
  {Rodda}, {Nelson}, {Nazarov}, {Mishevskiy}, {Myers}, {Denisenko}, {Stanek},
  {Shields}, {Kochanek}, {Holoien}, {Shappee}, {Prieto}, {Itagaki},
  {Nishiyama}, {Kabashima}, {Stubbings}, {Schmeer}, {Muyllaert}, {Horie},
  {Shears}, {Poyner}, \& {Moriyama}}]{Kato17}
{Kato}, T., {Isogai}, K., {Hambsch}, F.-J., {et~al.} 2017, \pasj, 69, 75

\bibitem[{{Kato} {et~al.}(2012){Kato}, {Maehara}, {Miller}, {Ohshima},
  {Miguel}, {Tanabe}, {Imamura}, {Akazawa}, {Kunitomi}, {Takagi}, {Nose},
  {Hambsch}, {Kiyota}, {Pavlenko}, {Baklanov}, {Antonyuk}, {Samsonov},
  {Sosnovskij}, {Antonyuk}, {Andreev}, {Morelle}, {Dubovsky}, {Kudzej},
  {Oksanen}, {Masi}, {Krajci}, {Pickard}, {Sabo}, {Itoh}, {Stein}, {Dvorak},
  {Henden}, {Nakagawa}, {Noguchi}, {Iino}, {Matsumoto}, {Nishitani}, {Aoki},
  {Kobayashi}, {Akasaka}, {Bolt}, {Shears}, {Ruiz}, {Shugarov}, {Chochol},
  {Parakhin}, {Monard}, {Shiokawa}, {Kasai}, {Staels}, {Miyashita}, {Starkey},
  {{\"O}gmen}, {Littlefield}, {Katysheva}, {Sergey}, {Denisenko}, {Tordai},
  {Fidrich}, {Goranskij}, {Virtanen}, {Crawford}, {Pietz}, {Koff}, {Boyd},
  {Brady}, {James}, {Goff}, {Itagaki}, {Nishimura}, {Nakashima}, {Yoshida},
  {Stubbings}, {Poyner}, {Maeda}, {Korotkiy}, {Sokolovsky}, \&
  {Ueda}}]{Kato2012}
{Kato}, T., {Maehara}, H., {Miller}, I., {et~al.} 2012, \pasj, 64, 21

\bibitem[{{Knigge} {et~al.}(2011){Knigge}, {Baraffe}, \&
  {Patterson}}]{Knigge11}
{Knigge}, C., {Baraffe}, I., \& {Patterson}, J. 2011, \apjs, 194, 28

\bibitem[{{Kolb} \& {Baraffe}(1999)}]{KolbBaraffe99}
{Kolb}, U. \& {Baraffe}, I. 1999, \mnras, 309, 1034

\bibitem[{{Kornet} \& {Rozyczka}(2000)}]{KornetRozyczka}
{Kornet}, K. \& {Rozyczka}, M. 2000, \actaa, 50, 163

\bibitem[{{Lasota}(2001)}]{Lasota2001}
{Lasota}, J.-P. 2001, \nar, 45, 449

\bibitem[{{Marsh} {et~al.}(1990){Marsh}, {Horne}, {Schlegel}, {Honeycutt}, \&
  {Kaitchuck}}]{MarshUGem}
{Marsh}, T.~R., {Horne}, K., {Schlegel}, E.~M., {Honeycutt}, R.~K., \&
  {Kaitchuck}, R.~H. 1990, \apj, 364, 637

\bibitem[{{Marsh} \& {Schwope}(2016)}]{MarshSchwope16}
{Marsh}, T.~R. \& {Schwope}, A.~D. 2016, Astrophysics and Space Science
  Library, Vol. 439, {Doppler Tomography}, ed. H.~M.~J. {Boffin}, G.~{Hussain},
  J.-P. {Berger}, \& L.~{Schmidtobreick}, 195

\bibitem[{{Mason} {et~al.}(2000){Mason}, {Skidmore}, {Howell}, {Ciardi},
  {Littlefair}, \& {Dhillon}}]{WZ2}
{Mason}, E., {Skidmore}, W., {Howell}, S.~B., {et~al.} 2000, \mnras, 318, 440

\bibitem[{{Morales-Rueda} \& {Marsh}(2002)}]{MoralesRueda2002}
{Morales-Rueda}, L. \& {Marsh}, T.~R. 2002, \mnras, 332, 814

\bibitem[{{Neustroev} {et~al.}(2019{\natexlab{a}}){Neustroev}, {Boyd},
  {Berardi}, {Zharikov}, {Medina}, {Page}, {Osborne}, {Kuin}, {Knigge},
  {Marsh}, {Gaensicke}, {Franco}, {Teyssier}, \& {Sjoberg}}]{NeustroevTCP1}
{Neustroev}, V., {Boyd}, D., {Berardi}, P., {et~al.} 2019{\natexlab{a}}, The
  Astronomer's Telegram, 13009

\bibitem[{{Neustroev} {et~al.}(2019{\natexlab{b}}){Neustroev}, {Watkins},
  {Kvist}, {Halsio}, {Ruokanen}, {Anetjarvi}, {Tordai}, {Page}, {Osborne},
  {Sjoberg}, {Boyd}, {Marsh}, {Gaensicke}, {Knigge}, {Zharikov}, {Rautio},
  {Rikkola}, {Poranen}, {Sarkar}, \& {Kuin}}]{NeustroevTCP2}
{Neustroev}, V., {Watkins}, A.~E., {Kvist}, P.~E., {et~al.} 2019{\natexlab{b}},
  The Astronomer's Telegram, 13297, 1

\bibitem[{{Neustroev} {et~al.}(2017){Neustroev}, {Marsh}, {Zharikov}, {Knigge},
  {Kuulkers}, {Osborne}, {Page}, {Steeghs}, {Suleimanov}, {Tovmassian},
  {Breedt}, {Frebel}, {Garc{\'{\i}}a-D{\'{\i}}az}, {Hambsch}, {Jacobson},
  {Parsons}, {Ryu}, {Sabin}, {Sjoberg}, {Miroshnichenko}, {Reichart},
  {Haislip}, {Ivarsen}, {LaCluyze}, \& {Moore}}]{NeustroevSSS}
{Neustroev}, V.~V., {Marsh}, T.~R., {Zharikov}, S.~V., {et~al.} 2017, \mnras,
  467, 597

\bibitem[{{Neustroev} {et~al.}(2011){Neustroev}, {Suleimanov}, {Borisov},
  {Belyakov}, \& {Shearer}}]{NeustroevUX}
{Neustroev}, V.~V., {Suleimanov}, V.~F., {Borisov}, N.~V., {Belyakov}, K.~V.,
  \& {Shearer}, A. 2011, \mnras, 410, 963

\bibitem[{{Neustroev} {et~al.}(2006){Neustroev}, {Zharikov}, \&
  {Michel}}]{NeustroevBZ}
{Neustroev}, V.~V., {Zharikov}, S., \& {Michel}, R. 2006, \mnras, 369, 369

\bibitem[{{Neustroev} {et~al.}(2016){Neustroev}, {Zharikov}, \&
  {Borisov}}]{Paper1}
{Neustroev}, V.~V., {Zharikov}, S.~V., \& {Borisov}, N.~V. 2016, \aap, 586, A10
  (Paper I)

\bibitem[{{Osaki}(1989)}]{Osaki89}
{Osaki}, Y. 1989, \pasj, 41, 1005

\bibitem[{{Osaki}(1996)}]{Osaki96}
{Osaki}, Y. 1996, \pasp, 108, 39

\bibitem[{{Osaki}(2005)}]{Osaki05}
{Osaki}, Y. 2005, Proceeding of the Japan Academy, Series B, 81, 291

\bibitem[{{Paczynski}(1977)}]{Paczynski}
{Paczynski}, B. 1977, \apj, 216, 822

\bibitem[{{Pala} {et~al.}(2020){Pala}, {G{\"a}nsicke}, {Breedt}, {Knigge},
  {Hermes}, {Gentile Fusillo}, {Holland s}, {Naylor}, {Pelisoli}, {Schreiber},
  {Toonen}, {Aungwerojwit}, {Cukanovaite}, {Dennihy}, {Manser}, {Pretorius},
  {Scaringi}, \& {Toloza}}]{Pala19}
{Pala}, A.~F., {G{\"a}nsicke}, B.~T., {Breedt}, E., {et~al.} 2020, \mnras, 494,
  3799

\bibitem[{{Pala} {et~al.}(2017){Pala}, {G{\"a}nsicke}, {Townsley}, {Boyd},
  {Cook}, {De Martino}, {Godon}, {Haislip}, {Henden}, {Hubeny}, {Ivarsen},
  {Kafka}, {Knigge}, {LaCluyze}, {Long}, {Marsh}, {Monard}, {Moore}, {Myers},
  {Nelson}, {Nogami}, {Oksanen}, {Pickard}, {Poyner}, {Reichart}, {Rodriguez
  Perez}, {Schreiber}, {Shears}, {Sion}, {Stubbings}, {Szkody}, \&
  {Zorotovic}}]{Pala17}
{Pala}, A.~F., {G{\"a}nsicke}, B.~T., {Townsley}, D., {et~al.} 2017, \mnras,
  466, 2855

\bibitem[{{Papaloizou} \& {Pringle}(1977)}]{PapaloizouPringle77}
{Papaloizou}, J. \& {Pringle}, J.~E. 1977, \mnras, 181, 441

\bibitem[{{Patterson}(1981)}]{Patterson81}
{Patterson}, J. 1981, \apjs, 45, 517

\bibitem[{{Skidmore} {et~al.}(2000){Skidmore}, {Mason}, {Howell}, {Ciardi},
  {Littlefair}, \& {Dhillon}}]{WZ1}
{Skidmore}, W., {Mason}, E., {Howell}, S.~B., {et~al.} 2000, \mnras, 318, 429

\bibitem[{{Smak}(1969)}]{Smak1969}
{Smak}, J. 1969, \actaa, 19, 155

\bibitem[{{Smak}(1981)}]{Smak1981}
{Smak}, J. 1981, \actaa, 31, 395

\bibitem[{{Smak}(1984)}]{Smak84}
{Smak}, J. 1984, \pasp, 96, 5

\bibitem[{{Smak}(1996)}]{Smak96}
{Smak}, J. 1996, \actaa, 46, 377

\bibitem[{{Smith} {et~al.}(2007){Smith}, {Haswell}, {Murray}, {Truss}, \&
  {Foulkes}}]{Smith07}
{Smith}, A.~J., {Haswell}, C.~A., {Murray}, J.~R., {Truss}, M.~R., \&
  {Foulkes}, S.~B. 2007, \mnras, 378, 785

\bibitem[{{Spruit} \& {Taam}(2001)}]{SpruitTaam01}
{Spruit}, H.~C. \& {Taam}, R.~E. 2001, \apj, 548, 900

\bibitem[{{Steeghs} {et~al.}(1997){Steeghs}, {Harlaftis}, \&
  {Horne}}]{Steeghs:1997aa}
{Steeghs}, D., {Harlaftis}, E.~T., \& {Horne}, K. 1997, \mnras, 290, L28

\bibitem[{{Steeghs} \& {Stehle}(1999)}]{SteeghsStehle}
{Steeghs}, D. \& {Stehle}, R. 1999, \mnras, 307, 99

\bibitem[{{Subebekova} {et~al.}(2020){Subebekova}, {Zharikov}, {Tovmassian},
  {Neustroev}, {Wolf}, {Hernandez}, {Ku{\v{c}}{\'a}kov{\'a}}, \&
  {Khokhlov}}]{RWTri}
{Subebekova}, G., {Zharikov}, S., {Tovmassian}, G., {et~al.} 2020, \mnras, 497,
  1475

\bibitem[{{Taam} {et~al.}(2003){Taam}, {Sandquist}, \&
  {Dubus}}]{TaamSandquistDubus03}
{Taam}, R.~E., {Sandquist}, E.~L., \& {Dubus}, G. 2003, \apj, 592, 1124

\bibitem[{{Taam} \& {Spruit}(2001)}]{TaamSpruit01}
{Taam}, R.~E. \& {Spruit}, H.~C. 2001, \apj, 561, 329

\bibitem[{{Teyssier}(2019)}]{Teyssier19}
{Teyssier}, F. 2019, The Astronomer's Telegram, 12936

\bibitem[{{Truss}(2007)}]{Truss}
{Truss}, M.~R. 2007, \mnras, 376, 89

\bibitem[{{van Spaandonk} {et~al.}(2010){van Spaandonk}, {Steeghs}, {Marsh}, \&
  {Torres}}]{GW_Lib}
{van Spaandonk}, L., {Steeghs}, D., {Marsh}, T.~R., \& {Torres}, M.~A.~P. 2010,
  \mnras, 401, 1857

\bibitem[{{Warner}(1995)}]{Warner95}
{Warner}, B. 1995, Cambridge Astrophysics Series, 28

\bibitem[{{Whitehurst} \& {King}(1991)}]{WhitehurstKing91}
{Whitehurst}, R. \& {King}, A. 1991, \mnras, 249, 25

\bibitem[{{Willems} {et~al.}(2005){Willems}, {Kolb}, {Sandquist}, {Taam}, \&
  {Dubus}}]{Willems05}
{Willems}, B., {Kolb}, U., {Sandquist}, E.~L., {Taam}, R.~E., \& {Dubus}, G.
  2005, \apj, 635, 1263

\end{thebibliography}
\end{document}